\documentclass[11pt,fleqn]{article}

\usepackage{amsmath,amsfonts,amssymb}

\def \lsim{\mathrel{\vcenter
{\hbox{$<$}\nointerlineskip\hbox{$\sim$}}}}
\def \gsim{\mathrel{\vcenter
{\hbox{$>$}\nointerlineskip\hbox{$\sim$}}}}

\def\bari{i \hspace{-3pt} \bar{\;\raisebox{3.7pt}{}}}
\def\barj{j \hspace{-3pt} \bar{\;\raisebox{3.7pt}{}}}

\def\bea{\begin{eqnarray}}
\def\eea{\end{eqnarray}}
\def\be{\begin{equation}}
\def\ee{\end{equation}}
\def\ba{\begin{array}}
\def\ea{\end{array}}
\def\nn{\nonumber}
\def\a{& \hspace{-11pt}}
\def\summ{\mbox{\large ${\sum}$}}

\font\tenrsfs=rsfs10
\font\sevenrsfs=rsfs7
\font\fiversfs=rsfs5
\newfam\rsfsfam
\textfont\rsfsfam=\tenrsfs
\scriptfont\rsfsfam=\sevenrsfs
\scriptscriptfont\rsfsfam=\fiversfs
\def\mathscr#1{{\fam\rsfsfam\relax#1}}

\begin{document}

\thispagestyle{empty}

\begin{center}

$\;$

\vspace{1.5cm}

{\huge \bf Metastable supergravity vacua with \\[3mm] $F$ and $D$ supersymmetry breaking}

\vspace{1.5cm}

{\Large {\bf Marta~G\'omez--Reino} \hspace{2pt} and \hspace{2pt} {\bf Claudio~A.~Scrucca}} \\[2mm] 

\vspace{1cm}

{\large \em Institut de Physique, Universit\'e de Neuch\^atel,\\ 
Rue Breguet 1, CH-2000 Neuch\^atel, Switzerland}
\vspace{.2cm}

\end{center}

\vspace{1cm}

\centerline{\bf \large Abstract}
\begin{quote}

We study the conditions under which a generic supergravity model involving chiral and 
vector multiplets can admit viable metastable vacua with spontaneously broken 
supersymmetry and realistic cosmological constant. To do so, we impose that on the 
vacuum the scalar potential and all its first derivatives vanish, and derive a necessary 
condition for the matrix of its second derivatives to be positive definite. 
We study then the constraints set by the combination of the flatness condition needed for 
the tuning of the cosmological constant and the stability condition that is necessary to 
avoid unstable modes. We find that the existence of such a viable vacuum implies a 
condition involving the curvature tensor for the scalar geometry and the charge and mass 
matrices for the vector fields. Moreover, for given curvature, charges and masses satisfying 
this constraint, the vector of $F$ and $D$ auxiliary fields defining the Goldstino direction is 
constrained to lie within a certain domain. The effect of vector multiplets relative to chiral 
multiplets is maximal when the masses of the vector fields are comparable to the gravitino mass. 
When the masses are instead much larger or much smaller than the gravitino mass, the effect becomes 
small and translates into a correction to 
the effective curvature. We finally apply our results to some simple classes 
of examples, to illustrate their relevance.

\vspace{5pt}
\end{quote}

\newpage

\renewcommand{\theequation}{\thesection.\arabic{equation}}

\section{Introduction}
\setcounter{equation}{0}

Recently substantial progress has been achieved in understanding how spontaneous 
supersymmetry breaking could be realized in a phenomenologically and cosmologically 
viable way in string inspired supergravity models. On one hand, from the microscopic string 
point of view, the structure of the K\"ahler potential is well understood at leading order in the 
weak coupling and low energy expansions \cite{logK,kounnas}, and also 
the effects leading to non-trivial superpotentials are now believed to be reasonably well 
understood \cite{GC,GVW,GKP}, although the identification of viable models still remains 
an open problem. Actually, it has been suggested that perhaps one should not look for a 
unique candidate that would be singled out for some reason, but rather for a statistical 
distribution in the landscape of possible vacua \cite{Stat}.
On the other hand, from the macroscopic supergravity point of view, much progress has 
been made in understanding the structure that the K\"ahler potential and the superpotential 
need to have in order for the theory to admit a phenomenologically acceptable vacuum. 
It has also been argued that the tuning of the cosmological constant could be realized in 
a very economical and transparent way in a theory including two sectors, one of them 
admitting an AdS supersymmetric vacuum and the other one breaking supersymmetry and 
adding a positive contribution to the potential \cite{KKLT}.

Spontaneous supersymmetry breaking can be triggered both by the $F$ auxiliary fields of 
chiral multiplets and the $D$ auxiliary fields of vector multiplets. It is however believed that 
the qualitative seed for the breaking must come from chiral multiplets, and that vector 
multiplets may only affect the quantitative aspects of it. The reason for this is that in 
standard situations the values of the $D$'s turn out to be proportional to the values of the 
$F$'s at any stationary point of the superpotential. This is true both in rigid \cite{ADM} and 
local supersymmetry \cite{KawamuraDFF}, and relies essentially on the holomorphicity of the 
superpotential. 
The only situation where this relation can be possibly avoided is in the presence of a genuine 
field-independent Fayet-Iliopoulos term. However, although in global supersymmetry this is 
a natural possibility, in local supersymmetry it is highly constrained due to the fact that the 
usual Fayet-Iliopoulos term is not invariant in the presence of gravity \cite{noFIsugra}. 
In fact, it turns out that even in the presence of a constant Fayet-Iliopoulos term associated 
to a gauged $R$ symmetry, the $D$'s are still proportional to the $F$'s, as long as the 
superpotential does not vanish \cite{FIsugra}, as is required in order to achieve a finite 
supersymmetry breaking scale with vanishing cosmological constant.
%The only situation where this relation can be avoided is in the presence of a 
%genuine field-independent Fayet-Iliopoulos term. However, although in global supersymmetry 
%this is a natural possibility, in local supersymmetry it can happen only in the very peculiar case 
%of a gauged $R$ symmetry \cite{noFIsugra}, which does not seem to naturally emerge from 
%any string construction \footnote{See \cite{FIsugra} for a more extensive discussion on this issue.}. 
%We will then concentrate on models where this exceptional possibility is not realized. 
Nevertheless, despite this relation between the $F$ and $D$ auxiliary fields, there are
scenarios in which the $D$-terms may play a crucial r\^ole besides the $F$-terms in 
producing a viable vacuum.

The situation in models where supersymmetry breaking is dominated by chiral multiplets
is by now well understood. It is known that the form of the K\"ahler potential imposes crucial 
restrictions on the possibility of getting a viable vacuum, that is, a metastable stationary point 
with a very tiny cosmological constant. It was for example argued in several ways that just the dilaton 
modulus could not lead to a viable situation \cite{nodildom}, unless subleading corrections 
to its K\"ahler potential become large \cite{nonpertdil,yetmoredil}, and therefore the perturbative
control over the theory is compromised. Similarly, it was realized that the compactification 
volume modulus could instead dominate the supersymmetry breaking, but only if subleading 
corrections to its K\"ahler potential are taken into account \cite{nonpertvol,nonpertvolbis}. 
In two recent papers \cite{grs1,grs2}, we have addressed in a more systematic way the 
general question of whether it is possible to translate more directly onto the parameters of 
a given theory the condition of flatness, related to the cancellation of the cosmological 
constant, and the condition of stability, related to the stabilization of all the fluctuation modes. 
We found that, in the simplest case of supergravity models involving only chiral multiplets, it 
is possible to answer to this question in a remarkably sharp and simple way. It turns out that 
there exists a strong necessary condition on the value that the curvature of the K\"ahler 
geometry is allowed to take. Moreover, the Goldstino vector of $F$ auxiliary fields gets 
constrained not only on its length, but also on its direction. More precisely, the form of the 
Riemann tensor must present valleys where the curvature takes a value that is below a 
certain threshold, and the Goldstino direction cannot  be too far away from those directions, 
how far this can be being fixed by the value of the curvature scalars. 

The results of \cite{grs1,grs2} provide a criterium that allows to discriminate to some extent 
between promising and non-promising models by knowing the approximate form of their 
K\"ahler potential only, independently of the form of the superpotential. Indeed, they were 
shown to have very useful and general implications on the possibility of realizing in a viable 
way supersymmetry breaking when it is dominated by the $F$ auxiliary fields. 
These implications are particularly striking for string models, where the moduli 
sector is identified with the hidden sector. The main reason for this is that in these 
models the K\"ahler geometry has constant curvature at leading order in the coupling and 
derivative expansions, so that the constraints translate very directly into restrictions on the 
parameters of the Lagrangian. For instance, the lower bound found for the curvature was 
used to explain in a more robust way why the dilaton cannot dominate supersymmetry 
breaking, whereas the volume modulus can dominate it but in a way that is very sensitive 
to small subleading corrections to the K\"ahler potential. A similar strategy has also been 
used to explore the statistics of supersymmetry breaking vacua in certain classes of string 
models \cite{Denef}. Finally, there have also been studies adopting a complementary 
viewpoint and developing more efficient tools to study algebraically the vacua allowed by 
a given theory with fixed K\"ahler potential and superpotential \cite{Lukas}.

The situation in models where supersymmetry breaking is significantly affected not only by 
chiral multiplets, but also by vector multiplets, is more complex and less understood. 
The fact that the $D$-term contribution to the scalar potential is positive definite, (unlike the 
$F$-term contribution, which has an indefinite sign), suggests that they may play an 
important r\^ole in the stabilization of the scalar fields triggering spontaneous supersymmetry 
breaking. Note that this distinction is essentially implied by Lorentz invariance, which allows 
scalar fields to have a generic potential leading to non-vanishing VEVs, but requires vector 
fields to have a restricted potential leading to vanishing VEVs. As a result, the squared masses 
of all the charged fields tend to be increased by the gauging. Actually there are two types of 
effects. The first is that Goldstone fields corresponding to global symmetries of the ungauged 
theory get absorbed by gauge fields through the Higgs mechanism when the coupling is switched 
on, and thereby obtain a positive physical squared mass. The second is that the non-Goldstone 
fields also receive a positive correction to their squared masses due to the gauging, which can 
therefore help in stabilizing them. These possibilities have been recently explored for instance 
in \cite{ADM,vz1,vz2,decarlos,lalak,dudasmambrini,ss,fp}, but a more systematic and quantitative 
understanding of the features of this more general situation is still missing. The aim of this work is 
to try to fill this gap by performing a general study from which one could extract information that 
can be directly used in practice to build viable models.

The main goal of this paper is thus to generalize the study performed in \cite{grs1,grs2} to the 
more general class of supergravity theories involving not only chiral multiplets but also vector 
multiplets that gauge isometries of the K\"ahler geometry. This can actually be done by following 
the same strategy as in \cite{grs1,grs2}. Conceptually, the main novelties lie in the fact that besides 
the K\"ahler potential and the superpotential, one must also specify the Killing vectors defining the 
isometries that are gauged and the gauge kinetic function defining the couplings. Technically the 
task is substantially more complicated due to the fact that the Goldstino vector now involves not 
only $F$ auxiliary fields but also $D$ auxiliary fields, and also that the scalar potential has a more 
complicated structure. Nevertheless, it is possible to find the exact generalization of the flatness 
and stability conditions of \cite{grs1}, but their implications can be worked out in detail only in certain 
specific regimes, due to the increased algebraic complication of the problem. In general, the effect 
of vector multiplets alleviates the necessary conditions for flatness and stability, compared to the 
case involving only chiral multiplets. However, the associated gauge symmetries also restrict the 
variety of models, as the superpotential should be gauge invariant up to K\"ahler transformations. 
We will actually show that in a variety of situations, the net qualitative effect of vector multiplets is 
to reduce the effective curvature felt by the chiral multiplets. In certain situations, this can be more 
sharply interpreted as coming from a correction to the effective K\"ahler potential induced by the 
presence of vector multiplets. As already mentioned, corrections to the K\"ahler potential can be 
crucial for the existence of a viable vacuum in certain situations. In this respect, there is a very 
interesting distinction between corrections induced by extra chiral multiplets and those induced 
by vector multiplets: the former can either increase or decrease the curvature, since they have an 
indefinite sign, whereas the latter (as we will show in the body of the paper) always decrease it.

A final comment is in order regarding the issue of implementing the idea of getting a metastable 
Minkowski minimum through an uplifting
sector that breaks supersymmetry in a soft way. It is clear that such a sector will have to contain
some light degrees of freedom, providing also some non-vanishing $F$ and/or $D$ auxiliary field.
Models realizing an $F$-term uplifting are easy to construct. A basic precursor of such models 
was first constructed in \cite{lutysundrum} and then further exploited, for instance, in \cite{Noiup}. 
More recently, a variety of other examples have been constructed, where the extra chiral multiplets
have an O' Raifeartaigh like dynamics, which is either genuinely postulated from the beginning 
\cite{FupNilles,Fup} or effectively derived from the dual description of a strongly coupled theory \cite{Fupiss} 
admitting a metastable supersymmetry breaking vacuum as in \cite{ISS}. Actually, a very simple 
and general class of such models can be constructed by using as uplifting sector any kind of sector 
breaking supersymmetry at a scale much lower than the Planck scale \cite{grs1}. Models realizing 
a $D$-term uplifting, on the other hand, are difficult to achieve. The natural idea of relying on some 
Fayet-Iliopoulos term \cite{BKQ} does not work, due to the already mentioned fact that such terms 
must generically be field-dependent in supergravity, so that the induced $D$ is actually 
proportional to the available charged $F$'s. It is then clear that there is an obstruction in getting 
$D$ much bigger than the $F$'s. Most importantly, if the only charged chiral multiplet in the model 
is the one of the would-be supersymmetric sector (which is supposed to have vanishing $F$) then 
also $D$ must vanish, implying that a vector multiplet cannot act alone as an uplifting sector
\cite{ChoiDup,DealwisDup}. This difference between $F$-term and $D$-term uplifting is once again 
due to the basic fact that chiral multiplets can dominate supersymmetry breaking whereas vector 
multiplets cannot.

The paper is organized as follows: 
In section 2, we present a brief summary of the relevant features of gauge invariant supergravity models. 
In section 3 we discuss the general properties of the vacuum and the interplay between the spontaneous 
breaking of supersymmetry and gauge symmetries. 
In section 4 we study various types of relations that exist between $D$ and $F$ auxiliary fields. 
In section 5 we derive a necessary condition for stability, and combine it with the flatness condition
to define a set of general constraints that are necessary for the existence of a viable vacuum.
In section 6 we elaborate on the possible strategies that can be used to derive more concrete implications 
from these constraints, and in particular on the different ways in which the relation between $D$ and $F$ 
auxiliary fields can be taken into account. 
In sections 7, 8 and 9 we then pursue three different approaches 
to this problem, which are based respectively on a dynamical relation, a kinematical relation and a 
kinematical bound between the $D$ and the $F$ auxiliary fields. 
In section 10 we illustrate the relevance of our general results with some examples of string 
inspired models.
In section 11, we conclude with a qualitative summary of our results.
 
\section{Gauge invariant supergravity models}
\setcounter{equation}{0}

In this section we will briefly review the main features of general supergravity models 
with minimal supersymmetry in four dimensions \cite{sugra1,sugra2}, emphasizing those
particular aspects that will be relevant for our analysis. We will use the notation of 
\cite{WessBagger}  and set $M_{\rm P}=1$.

Consider first a supergravity theory with $n$ chiral multiplets $\Phi^i$. The two-derivative 
Lagrangian is specified by a single real K\"ahler function $G(\Phi^k,\Phi^{k \dagger})$ 
\footnote{The function $G$ is related to the more commonly used K\"ahler potential $K$
and superpotential $W$ by the equation $G = K + {\rm log} |W|^2$. This decomposition
is however ambiguous, due to the K\"ahler symmetry transforming $K \to K + F + \bar F$
and $W \to e^{-F}W$, and leaving $G$ invariant. As already mentioned, $W$ cannot 
vanish at a non-supersymmetric Minkowski minimum, and the function $G$ is therefore
well defined.}. 
The K\"ahler geometry of the manifold spanned by the complex scalar fields is determined
by the metric $g_{i \barj} = G_{i\barj}$, which can be used to raise and lower chiral indices
\footnote{Subscripts on scalar quantities denote ordinary derivatives with respect to the fields.}. 
It can happen that this theory has a group of some number $m$ of global symmetries, 
compatibly with supersymmetry. These are generated by holomophic Killing vectors 
$X_a^i(\Phi^k)$, in the sense that a generic symmetry transformation with infinitesimal 
real constant parameters $\lambda^a$ is implemented by the operator
$\delta = \lambda^a (X_a^i \partial_i + X_a^{\bari} \partial_{\bari})$. The chiral superfields 
transform as $\delta \Phi^i = \lambda^a X_a^i(\Phi^k)$. It is then clear that the condition 
for these transformations to represent an invariance of the theory is that the function $G$ 
should be invariant: $\delta G = 0$. 
This implies:
\be
X_a^i G_i + X_a^{\bari} G_{\bari} = 0 \,.
\label{Ginv}
\ee
This equation is valid at any point of the scalar manifold and one can therefore 
derive additional conditions by taking derivatives of it. The corresponding information 
is most conveniently extracted by using covariant derivatives $\nabla_i$, in terms of which 
holomorphicity of the vectors $X_a^i$ implies $\nabla_i X_a^{\barj} = 0$ or $\nabla_i X_{aj} = 0$. 
Taking one derivative of eq.~(\ref{Ginv}), one finds that:
\be
X_{ai} + X_a^k \nabla_i G_k + G_k \nabla_i X_a^k = 0 \,.
\label{G1inv}
\ee
Taking two derivatives, and using the fact that the metric is covariantly constant, one finds 
instead:
\be
\nabla_i X_{a\barj} + \nabla_{\barj} X_{ai} = 0 \,.
\label{G2inv}
\ee
Using the definition of the Riemann tensor and the holomorphicity of the Killing vectors we know that 
$\nabla_i \nabla_{\barj} X_{ap} = R_{i \barj p \bar q} X_a^{\bar q}$. With the help of this relation, 
and taking a suitable combination of four derivatives of (\ref{Ginv}), one also finds:
\bea
\a\a \Big(\hspace{-1pt} X_a^m \nabla_{\hspace{-1pt} m} 
\!+\! X_a^{\bar n} \nabla_{\hspace{-1pt} \bar n} \hspace{-1pt} \Big) R_{i \barj p \bar q} 
\hspace{-1pt} +\! R_{r \barj p \bar q}  \nabla_{\hspace{-1pt} i}  X_a^r  
\hspace{-1pt} +\! R_{i \bar s p \bar q} \nabla_{\hspace{-1pt} \barj} X_a^{\bar s} 
\hspace{-1pt} +\! R_{i \barj t \bar q} \nabla_{\hspace{-1pt} p} X_a^t 
\hspace{-1pt} +\! R_{i \barj p \bar u} \nabla_{\hspace{-1pt} \bar q} X_a^{\bar u} 
= 0\,. \hspace{-100pt}
\label{G4inv}
\eea
The conditions (\ref{G2inv}) and (\ref{G4inv}) can be rephrased in terms of Lie derivatives 
as ${\cal L}_{X_a} \, g_{i \barj} = 0$ and ${\cal L}_{X_a} \, R_{i \barj p \bar q} = 0$, and 
show that each symmetry is associated to an isometry of the scalar manifold  
\footnote{Notice that for constant curvature K\"ahler manifolds, for which $\nabla_{k} R_{i \barj p \bar q} = 0$, 
the conditions (\ref{G4inv}) represent linear constraints on the derivatives of the Killing vectors.}.

Consider now the possibility of gauging such isometries with the introduction of vector multiplets. 
The corresponding supergravity theory will then include $n$ chiral multiplets $\Phi^i$ and $m$ 
vector multiplets $V^a$. Its two-derivative Lagrangian is specified by a real K\"ahler function 
$G(\Phi^k,\Phi^{k \dagger},V^a)$, determining in particular the scalar geometry, $m$ holomorphic 
Killing vectors $X_a^i(\Phi^k)$, generating the isometries that are gauged, and an $m$ by $m$ 
matrix of holomorphic gauge kinetic functions $H_{ab}(\Phi^k)$, defining the gauge couplings. 
There exists a general and systematic way of promoting the globally invariant action of the chiral 
multiplet theory to a locally invariant Lagrangian involving also the vector multiplets. This can be 
done in superfields by generalizing the real constant transformation parameters $\lambda^a$ 
to chiral superfield parameters $\Lambda^a$, and asking the transformation operator $\delta$ 
to act also on the vector fields, as 
$\delta = \Lambda^a X_a^i \partial_i + \bar \Lambda^a X_a^{\bari} \partial_{\bari} 
- i \big(\Lambda^a - \bar \Lambda^a) \partial_a$. The transformations of the chiral and vector multiplets 
read then $\delta \Phi^i = \Lambda^a X_a^i(\Phi^k)$ and 
$\delta V^a = - i (\Lambda^a - \bar \Lambda^a)$. 
The minimal coupling between chiral and vector multiplets turn ordinary derivatives into covariant 
derivatives, and induces a new contribution to the scalar potential coming from the vector auxiliary 
fields $D^a$, in addition to the standard one coming from the chiral auxiliary fields $F^i$. 
The condition for the action to be invariant under the just mentioned local transformations 
is clearly that $G$ must be invariant: $\delta G = 0$ \footnote{This means that the K\"ahler potential $K$ 
and the superpotential $W$ must be invariant only up to a local K\"ahler transformation,
associated with a gauging of the R symmetry.}. This implies now the conditions:
\be
G_a = - i \, X_a^i \, G_i = i \, X_a^{\bari} \, G_{\bari} \,.
\label{Grel}
\ee
In addition, the gauge kinetic function $H_{ab}$ must have an appropriate behavior under gauge 
transformations, in such a way as to cancel possible gauge anomalies $Q_{abc}$ and to lead to 
a consistent quantum effective action. More precisely, ${\rm Re}\,H_{ab}$ must be invariant, whereas 
${\rm Im}\,H_{ab}$ must have a variation that exactly matches the coefficient of $Q_{abc}$ 
\footnote{In general, there can also be generalized Chern-Simons terms, which can be added 
to the Lagrangian compatibly with supersymmetry. These can contribute to the anomalous variation 
of the action and allow therefore to generalize a bit the way in which anomaly cancellation is realized.
See \cite{AnoCSG} for discussions of this point. Here we shall simply interpret $Q_{abc}$ as the residual 
quantum anomaly that is left after having considered possible generalized Chern Simons terms.}.
In general, such anomalies have a quite involved structure and can depend on the scalar fields. 
Their general form has been studied in \cite{sigmaanomalies} in the limit 
of rigid supersymmetry, and more recently in \cite{sugraano} in the context of local supersymmetry. 
The functional form that $H_{ab}$ is allowed to take is then strongly constrained and linked to the form 
of the anomaly $Q_{abc}$. More precisely, the anomaly cancellation condition 
$\delta H_{bc} = i\, \lambda^a Q_{abc}$ implies the functional relations
\bea
X_a^i \nabla_i H_{bc} =  i \, Q_{abc} \,.
\label{ano}
\eea
Now, since the quantities $X_a^i$ and $H_{bc}$ are both holomorphic, these equations are 
subject to a strong integrability condition. Indeed, by taking first derivatives of the relations (\ref{ano}),
one finds that the real function $Q_{abc}$ must actually be constant. The functional form of the gauge 
kinetic function is then essentially 
fixed by the transformation rules of the scalar fields, in such a way that its variation is constant. 
This means that the possibility of canceling anomalies through local Wess-Zumino terms, which 
always exists in general, is quite strongly constrained by supersymmetry. In the low-energy effective
theories underlying string models, for instance, it is realized in the special form of the Green-Schwarz 
mechanism \cite{GS}. In that case, some fields that were neutral at tree-level and present 
in $H_{ab}$ acquire a non-trivial transformation law due to one-loop corrections to the 
K\"ahler potential. In addition, there can also be one-loop threshold corrections to the gauge kinetic 
function that depend on charged fields. These two effects account for anomaly cancellation in all 
the known situations.

The expectation values of the real and imaginary parts of $H_{ab}$ define respectively the inverse 
couplings and the $\theta$-angles for the vector fields. The former defines also a metric for the 
gauge fields, which can be used to raise and lower vector indices:
\be
h_{ab} = {\rm Re} H_{ab} \,.
\ee 
The expectation values of the first and second derivatives of $H_{ab}$ define additional relevant 
parameters of the theory. As a consequence of the holomophicity of $H_{ab}$, it is possible to 
parametrize these through the following two quantities:
\be
h_{abi} = \nabla_i {\rm Re} H_{ab} \,, \; \hspace{1cm}\;
h_{abij} = \nabla_i \nabla_j {\rm Re} H_{ab}\,.
\ee
The condition (\ref{ano}) for anomaly cancellation implies then the relation:
\bea
X_a^i h_{bc i} =  \frac i2 \, Q_{abc} \,.
\label{f0inv}
\eea

The theory is most conveniently formulated by using the superconformal formalism \cite{supconf}, 
with a chiral compensator multiplet $\Phi$. The Lagrangian reads then simply 
\footnote{The $D$ and $F$ densities contain now, besides the terms 
appearing in rigid supersymmetry, also terms depending on the graviton and the gravitino, 
but these will not be relevant for us.}:
\bea
\a\a {\cal L} = \int \!\! d^4 \theta \Big[ \!-\! 3\,\exp \Big\{\!\!-\!\frac 13 G(\Phi^k, \Phi^{k\dagger}, V^a) \Big\}\Big]\, 
\Phi^\dagger \Phi + \bigg(\int \!\! d^2 \theta \, \Phi^3 + {\rm h.c.} \bigg) \nn \\
\a\;\a \hspace{22pt} +\, \bigg( \int \!\! d^2 \theta \, \frac 14\,H_{ab}(\Phi^k) \, W^{a \alpha} W^b_{\alpha} 
+ {\rm h.c.} \bigg)\,.
\label{lag}
\eea
The kinetic Lagrangians for the scalar and the gauge fields involve respectively the covariant derivative 
${\cal D}_\mu \Phi^i= \partial_\mu \Phi^i - X_a^i (\Phi^k) A^a_\mu$ and the field strengths $F_{\mu\nu}^{a}$,
and read:
\be
{\cal L}_{\rm kin} = -\, g_{i \barj} \, {\cal D}^\mu \phi^i {\cal D}_\mu^* \phi^{\barj} 
- \frac 14 \, h_{ab} \, F^{\mu\nu a} F_{\mu\nu}^b \,.
\ee
The scalar potential can be computed by integrating out the auxiliary fields.  To do so, it is convenient 
to gauge fix the redundant superconformal symmetries by setting the scalar component of the compensator 
to $e^{G/6}$ and the fermion component to $0$, and to parametrize its non-trivial auxiliary component as 
$e^{G/6} F$. The auxiliary field Lagrangian is then easily computed, and the corresponding equations 
fix $F = e^{G/2} (1 - 1/3\, G^k G_k)$ and 
\be
F^i = - \, e^{G/2}\,g^{i \barj}\, G_\barj \,, \; \hspace{1cm}\;
D^a = -\,h^{ab}\, G_b \,. \label{FFiDa}
\ee
Substituting these expressions back into the auxiliary field Lagrangian, the scalar potential is found to be:
\be
V = e^G \Big(G^k G_k - 3 \Big) + \frac 12 G^a G_a \,.
\label{scalpot}
\ee
Using the relation (\ref{Grel}) and the expression $m_{3/2} = e^{G/2}$ for the gravitino mass, 
one can finally rewrite this in the standard way as  
\be
\label{genpot}
V = - 3\, m_{3/2}^2 + g^{i \barj}\, F_i F_{\barj} + \frac 12 h^{ab} D_a D_b\,.
\ee
where
\bea
\a\a F_i = - m_{3/2}\, G_i \,, \label{F} \\[1mm]
\a\a D_a = i \, X_a^i \, G_i = - i \, X_a^{\bari} \, G_{\bari} \,. \label{D}
\eea

The relation (\ref{D}) shows that the $D_a$ can be geometrically identified with the Killing 
potentials. Indeed, taking one derivative one finds that the Killing vectors can be written in 
the following way, which automatically solves the Killing equation (\ref{G2inv}):
\be
X_{ai} = i \, \nabla_i D_a \,, \; \hspace{1cm}\;
X_{a \bari} = - i\, \nabla_{\bari} D_a \,.
\label{XD}
\ee
Taking two derivatives of the relation (\ref{D}), one obtains instead a two-index tensor 
characterizing the local "charges" of the chiral multiplets
\be
q_{a i \barj} = \frac i2 \big(\nabla_i X_{a \barj} - \nabla_{\barj} X_{a i} \big) = \nabla_i \nabla_{\bar j} D_a \,.
\label{T}
\ee
Notice finally that by taking derivatives of (\ref{Grel}), one also deduces that 
$X_{ai} = - i\, G_{ai}$, $X_{a \bari} = i\, G_{a \bari}$ and $q_{a i \barj} = - G_{a i \barj}$.

\section{Vacuum and spontaneous symmetry breaking}
\setcounter{equation}{0}

The vacuum of the theory is associated to a stationary point of the scalar potential
(\ref{scalpot}). The scalar fields take in general non-vanishing vacuum expectation 
values, and local supersymmetry and the gauge symmetries can thus be spontaneously 
broken. The cosmological constant scale is identified with the vacuum energy, and we will 
assume that it is adjusted, through a tuning of parameters in the effective Lagrangian, to its 
very small measured value, which is negligible with respect to the gravitino mass scale 
$e^{G/2}$. We shall therefore impose from the beginning that on the vacuum $V=0$, implying the 
flatness condition:
\be
- 3 + G^i G_i + \frac 12 \, e^{-G} G^a G_a = 0\,.
\label{flatness}
\ee
The stationarity conditions correspond to requiring that $\nabla_i V = 0$, and they are given by:
\be
G_i + G^k \nabla_i G_k + e^{-G} \Big[G^a\Big(\nabla_i - \frac 12\, G_i \Big) G_a 
+ \frac 12\,  h_{abi} G^a G^b \Big] = 0\,.
\label{stationarity}
\ee

The $2n$-dimensional mass matrix for small fluctuations of the scalar fields around the 
vacuum has two different $n$-dimensional blocks, which can be computed as 
$m_{i \barj}^2 = \nabla_i \nabla_{\barj} V$ and $m_{i j}^2 = \nabla_i \nabla_j V$. Using 
the flatness and stationarity conditions, one finds, after a straightforward computation 
\cite{FKZ,dudasvempati}:
\bea
\a\a m_{i \barj}^2 = e^G \Big[g_{i \barj} - R_{i \barj p \bar q} G^p G^{\bar q} 
+  \nabla_i G_k \nabla_{\barj} G^k \Big] \nn \\
\a\;\a \hspace{30pt} \,+\, \Big[\!-\! \frac 12\, \Big(g_{i \barj} - G_i G_{\barj} \Big) G^a G_a
+ \Big(G_{(i} h_{ab \barj)} + h^{cd} h_{a c i} h_{b d \barj} \Big)\, G^a G^b \label{mijbar} \\[1mm]
\a\;\a \hspace{47pt} -\,2\, G^a G_{(i} \nabla_{\barj)} G_a - 2\, G^a h^{bc} h_{ab(i} \nabla_{\barj)} G_c
+ h^{ab} \nabla_i G_a  \nabla_{\barj} G_b + G^a \nabla_i \nabla_{\barj} G_a \Big] \,, \nn \\
\a\a m_{i j}^2 = e^G \Big[2 \, \nabla_{(i} G_{j)} + G^k \nabla_{(i} \nabla_{j)} G_k \Big] \nn \\
\a\;\a \hspace{30pt} \, +\,\Big[\!-\! \frac 12 \Big(\nabla_{(i} G_{j)} - G_i G_j \Big) G^a G_a
+ \Big(G_{(i} h_{abj)} + h^{cd} h_{a c i} h_{b d j} - \frac 12 h_{a b i j} \Big) G^a G^b  \nn \\[1mm]
\a\;\a \hspace{47pt} -\,2\, G^a G_{(i} \nabla_{j)}  G_a - 2\,G^a h^{bc} h_{ab(i} \nabla_{j)} G_c 
+ h^{ab}\, \nabla_i G_a \nabla_j G_b \Big] \,.
\label{mij}
\eea
The $m$-dimensional mass matrix for the vector fields can instead be read off from the kinetic 
term of the scalar fields, and has the form:
\be
M_{ab}^2 = 2\, g_{i \barj} X_a^i X_b^{\barj}
= 2\, g_{i \barj}\, \nabla^i G_a \nabla^{\barj} G_b \,.
\label{Mab}
\ee

In general the theory displays a spontaneous breakdown of both supersymmetry and gauge 
symmetries. These two breakings can happen at independent scales and along different directions 
in field space, and this is the main reason for the increased complexity of this analysis. Supersymmetry 
breaking is realized through a fermionic version of the Higgs mechanism, in which the gravitino field 
absorbs a linear combination of the fermion fields $\psi^i$ and $\psi^a$ of the chiral and vector
multiplets,  the would-be Goldstino fermion. The relevant combination can be read off from the mixing 
between the gravitino and the chiral and vector multiplet  fermions, and is given by:
\be
\eta = \frac {i}{\sqrt{3}} e^{G/2} G_i \psi^i + \frac {1}{\sqrt{6}} G_a \psi^a \,.
\ee
The Goldstino is thus associated with the vector $\eta_\xi = (i e^{G/2} G_i/\sqrt{3}, G_a/\sqrt{6})$
in the space of chiral and vector multiplet spinors $\psi^\xi=(\psi^i,\psi^a)$: $\eta = \eta_\xi \psi^\xi$. 
The norm squared of this vector is given by $\eta_\xi \eta^\xi = e^G G^i G_i/3 + G^a G_a/6 = m_{3/2}^2$,
due to the flatness condition (\ref{flatness}), and the scale of supersymmetry breaking can therefore 
be associated with the gravitino mass $m_{3/2}$. On the other hand, gauge symmetry breaking is 
realized through an ordinary bosonic Higgs mechanism, 
in which the gauge fields $A_\mu^a$ absorb linear combinations of the scalar fields $\phi^i$ of the 
chiral multiplets, the would-be Goldstone bosons. The relevant combinations can be read off from the 
mixing between the gauge fields and the chiral multiplet scalar fields, and are given by:
\bea
\sigma_a = X_{ai} \, \phi^i + X_{a\bari} \, \phi^{\bari}\,.
\eea
The Goldstone bosons are thus associated to the vectors $\sigma_{a \alpha} = (X_{ai},X_{a\bari})$
in the space of chiral multiplet scalars $\phi^\alpha = (\phi^i,\phi^{\bari})$: 
$\sigma_{a} = \sigma_{a\alpha} \phi^\alpha$. The scalar product between two of these vectors 
is given by $\sigma_{a\alpha} \sigma_b^\alpha = 2 X_a^i X_{bi} = M^2_{ab}$, and the scales 
of gauge symmetry breaking are therefore controlled by the gauge field mass matrix $M_{ab}$.

In general, supersymmetry and gauge symmetry breaking occur in an entangled way because 
of two reasons. The first is that the directions of the breakings are in general not orthogonal, and the 
second is that the breaking scales are not necessarily well separated. This situation simplifies 
(and the two types of breakings disentangle) only whenever the Goldstino and the Goldstone 
directions are nearly orthogonal or the gravitino and the gauge field masses are hierarchically 
different. In the first case, the two breaking disentangle just because they involve different sets 
of fields, whereas in the second they decouple because of the very large mass 
difference of the relevant fields.

\section{Chiral versus vector auxiliary fields}
\setcounter{equation}{0}

The chiral and vector auxiliary fields are given by $F_i = - e^{G/2} G_i$ and $D_a = - G_a$.
These relations are enforced by the equations of motion of the auxiliary fields, 
and are therefore true at any point of the scalar field space, and not only at the stationary 
points of the potential. These two kinds of auxiliary fields are however not completely independent 
from each other. There is a first relation between the $F_i$ and the $D_a$, which is of ''kinematical" 
nature, in the sense that it is satisfied as a functional relation valid at any point of the scalar field space. 
This first relation comes from (\ref{Grel}) and is a consequence of the gauge invariance of $G$. It reads:
\be
D_a = - i \, \frac {X_a^i}{m_{3/2}}  F_i = i \, \frac {X_a^{\bari}}{m_{3/2}}F_{\bari} \,.
\label{DFkin}
\ee
This relation shows that the $D_a$ are actually linear combinations of the $F_i$, with 
coefficients of order ${\cal O}(M_{aa}/m_{3/2})$. One can then derive a simple bound on 
the sizes that the $D_a$ can have relative to the $F_i$. Indeed, using the inequality 
$|a^i b_i| \le \sqrt{a^i a_i} \sqrt{b^j b_j}$, one deduces from the relation (\ref{DFkin}) that:
\be
|D_a| \le \frac 1{\sqrt{2}} \frac {M_{aa}}{m_{3/2}} \sqrt{F^i F_i} \,.
\label{DF}
\ee

There is also a second relation between the $F_i$ and the $D_a$, which is instead of 
''dynamical" nature, in the sense that it is valid only at the stationary points of the potential. 
It comes from considering a 
suitable linear combination of the stationarity conditions (\ref{stationarity}) along the direction 
$X_a^i$, that is, imposing stationarity with respect to those particular field variations 
that correspond to complexified gauge transformations: $X_a^i \nabla_i V= 0$.
This corresponds essentially to extract the information concerning stationarity of those scalar
fields that are absorbed by the gauge field in the Higgs mechanism. To derive this relation, 
one starts by contracting (\ref{stationarity}) with $X_a^i$. The result can be simplified by using 
eq.~(\ref{G1inv}). Using also the relations (\ref{f0inv}), (\ref{D}) and (\ref{Mab}) one gets:
\be
- \nabla_i X_{a\barj} \, F^i F^{\barj} 
- i\, \Big(F^i F_i - m_{3/2}^2 \Big) \, D_a - \frac i2\, M^2_{ab} \, D^b
+\, \frac i 2 Q_{abc} D^b D^c = 0 \,.
\ee
The real part of this equation is identically satisfied at any point. Indeed, the real part of the first 
term vanishes due to the Killing condition (\ref{G2inv}), whereas that of the last three terms 
vanishes trivially due to the fact that the quantities $D_a$, $M^2_{ab}$ and $Q_{abc}$ are real.
This just reflects the fact that the potential $V$ is invariant under gauge transformations at any 
point: ${\rm Re} (X_a^i \nabla_i V) = \delta V = 0$. The imaginary part of this equation is 
instead non-trivial, as $V$ is not invariant under imaginary gauge transformations. Using the definition 
(\ref{T}), we get the following quadratic relation between the $D_a$ and the $F_i$ 
\cite{KawamuraDFF,ChoiDFF} (see also \cite{KawaKobDFF}):
\be
q_{ai\barj} \, F^i F^{\barj}  
- \frac 12\,\Big[ M^2_{ab} + 2 \Big(F^i F_i - m_{3/2}^2 \Big) h_{ab} \Big]\, D^b
+\, \frac 12 \, Q_{abc} D^b D^c = 0 \,.
\label{DFF}
\ee

It is worth discussing what happens in the limit of rigid supersymmetry. In order to do so, it is important
to note that all the formulae that have been derived up to now in this section are actually valid for any 
value of the cosmological constant, and not only for the case where it vanishes. This means 
that the sizes of the auxiliary fields $F_i$ and $D_a$ are not necessarily limited by the flatness condition. 
We can then consider the limit of rigid supersymmetry, $m_{3/2} \rightarrow 0$  and $M_P \rightarrow \infty$, 
while keeping all the other quantities fixed. In that limit, the kinematical constraint (\ref{DF}) becomes trivial. 
This reflects the fact that in the rigid limit, the functional forms of the $F_i$ and the $D_a$ simplify and become 
independent of each other. On the other hand, the dynamical relation (\ref{DFF}) persists in the rigid limit, 
although it gets substantially simplified. Indeed, the second and third terms in the relation (\ref{DFF}) can be 
neglected, since by dimensional analysis they are suppressed by powers of $M_P$. Therefore we are 
left with: $q_{ai\barj} \, F^i F^{\barj} - 1/2\,M^2_{ab} D^b + 1/2 \, Q_{abc} D^b D^c = 0$, and we can conclude 
that the kinematical bound (\ref{DF}) is of gravitational origin, whereas the dynamical relation (\ref{DFF}) is of 
non-gravitational origin. In the light of this observation, it should be emphasized that the obstruction in getting a 
non-zero $D$ in situations with vanishing $F$ should be interpreted as a constraint coming from the 
dynamical relation (\ref{DFF}) rather than from the kinematical relation (\ref{DFkin}), since it is due to the holomorphicity 
of the superpotential and not to any particular feature associated to gravity \footnote{Nevertheless, it is curious 
that in the presence of an anomaly cancelled by a Wess-Zumino term, eq. (\ref{DFF}) does not seem 
to display automatically this obstruction and one apparently needs to invoke (\ref{DFkin}).}. It is clear that this 
obstruction also translates into a natural tendency for $D$ auxiliary fields to be smaller than $F$ auxiliary fields 
\footnote{See for instance \cite{Dbreaking} for a discussion on the possibility of achieving $D$ larger than $F$ 
in the simplest class of theories with global supersymmetry and linearly realized gauge symmetries.}.

We are going to discuss now two situations in which the relation between the $F_i$ and the $D_a$ 
simplifies. They correspond to the two different limits where the breaking of supersymmetry and gauge 
symmetries happen at well separated scales and therefore disentangle. 

\subsection{Heavy vector limit}

A first limit in which the situation substantially simplifies is when the gauge bosons masses are much 
bigger than the gravitino mass: $M_{ab} \gg m_{3/2}$. In this limit (and assuming that all the other 
quantities in the problem remain finite) one finds that the kinematical bound (\ref{DF}) becomes 
irrelevant, whereas the dynamical relation (\ref{DFF}) implies that the $D_a$ 
are small and given by 
\be
D^a \simeq 2 \, M^{-2 ab} q_{bi \barj} \, F^i F^{\barj} \,.
\label{DFFlarge}
\ee
More precisely, the flatness condition implies 
that $F_i \sim {\cal O}(m_{3/2})$, and the expression (\ref{DFFlarge}) has thus the form of a sum of 
terms that are ${\cal O}(m_{3/2}/M_{ab})$, with ${\cal O}(q_{ai\barj})$ coefficients. The simplification 
of the relation (\ref{DFF}) in this limit is due to the fact that gauge symmetries are broken at a much higher 
scale than supersymmetry. It is then a good approximation to integrate out the vector multiplets in a 
supersymmetric way at the scale defined by $M_{ab}$, and consider supersymmetry breaking 
within a low-energy effective supergravity theory involving only chiral multiplets. The $F_i$ represent 
then the dominant supersymmetry breaking effects that are directly induced by chiral multiplets, whereas 
the $D_a$ can now be understood as encoding the subleading corrections to supersymmetry breaking 
effects that are indirectly induced by virtual heavy vector fields \cite{ADM}. 

Note also that due to the kinematical relation (\ref{DFkin}), the linear combination of chiral auxiliary 
fields $X_a^i F_i$ (which corresponds to the vector auxiliary fields $D_a$) must be small. One gets 
then the $m$ restrictions $X_a^i F_i \simeq 0$ as a consequence of the fact that, in the limit considered 
here, $m$ of the $n$ chiral multiplets are absorbed through a supersymmetric Higgs mechanism, and 
only $n-m$ gauge-invariant combinations of chiral multiplets remain as light fields in the low energy 
effective action that is used to describe supersymmetry breaking effects. Note that this indirectly implies 
that the vector fields can be much heavier than the gravitino (without any other parameter blowing up) 
only if the number of chiral multiplets is higher than the number of vector multiplets: $n > m$.

It is possible to verify that one correctly reproduces (\ref{DFFlarge}) by integrating out explicitly 
the vector fields. This can be done by solving the vector multiplet equations of motion directly in 
superfields. In the limit we are considering terms involving supercovariant derivatives can be 
neglected, since on dimensional grounds they would give subleading contributions. This means 
that one can neglect the gauge kinetic term in (\ref{lag}) and the equations of motion of the vector 
superfields are simply:
\be
G_a (\Phi_i, \Phi_i^\dagger, V_a) \exp \Big\{\!\!-\!\frac 13 G(\Phi_i, \Phi_i^\dagger, V_a) \Big\} \,
\Phi^\dagger \Phi \simeq 0\,.
\label{Veq}
\ee
These represent algebraic equations involving the vector superfields $V_a$, the chiral superfields 
$\Phi_i$, and the compensator superfield $\Phi$. To solve them and integrate out the vector superfields,
one needs to choose a gauge for the superfield gauge transformations. A convenient choice is the 
super-unitary gauge, where the would-be Goldstone superfields are set to zero. At the linearized level, 
this means setting $G_a \simeq 0$ (the $\simeq$ sign being related to the fact that this is possible 
only in the exactly supersymmetric limit). One can then solve eq.~(\ref{Veq}) and express 
the $V^a$ in terms of the physical $\Phi^i$ orthogonal to $X_a^i$ and $\Phi$. Substituting the 
results back into the Lagrangian, one would then find the whole effective Lagrangian for the light 
chiral multiplets. Except for very special models, this is however difficult to carry out explicitly in 
practice, due to the fact that arbitrary powers of $V^a$ appear in the equation. As a consequence, 
it is in general not possible to solve (\ref{Veq}) exactly at the superfield level to determine the 
complete effective Lagrangian in closed form. One can however solve it approximately by 
assuming that the superfields $V^a$ are small, and expanding it in powers of $V^a$. At first 
order in the expansion, one finds that this is consistent if $G_a \simeq 0$, as is the case in the 
super-unitary gauge we are using. At second order, the equation for $V^a$ is linear and can be 
solved to determine the leading order effect of the vector multiplet. In doing so one finds the following 
expression:
\be
\Big[G_{ab} - \frac 13 G_a G_b \Big] V^b + G_{a} \simeq 0 \,.
\label{Deq}
\ee
Taking into account that the quantity $G_{ab} - G_a G_b/3$ has a vacuum expectation value given by 
$M_{ab}^2/2$, which is by assumption large, one finally finds:
\be
V^a \simeq - 2 M^{-2ab} G_b \,.
\label{Vlin}
\ee
In this equation, $V^a$ and $G_b$ are superfields, whereas $M^{-2ab}$ are numbers. Taking 
finally the $D$ component of this equation, and recalling that $G_{a i \barj} = - q_{a i \barj}$, one 
recovers eq.~(\ref{DFFlarge}). One can also use the result (\ref{Vlin}) to compute the leading 
order correction that is induced by the heavy vector multiplets in the effective action for the light 
chiral multiplets. It is given by $\Delta G \simeq - M^{-2ab} G_a G_b$ \footnote{The correction 
$D^a D_a/2$ to the scalar potential is not included in this result, and corresponds to a subleading 
effect coming from the vector field kinetic terms.}. Notice finally that the $F$ component of the 
gauge fixing condition $G_a \simeq 0$ reproduces instead the constraint $X_a^i F_i \simeq 0$, 
since $G_{ai} = i X_{ai}$.

\subsection{Light vector limit}

Another limit in which the situation also simplifies is when the gauge bosons masses are much smaller 
than the gravitino mass: $M_{ab} \ll m_{3/2}$. In this limit (and assuming that all the other quantities in 
the problem remain finite) the kinematical bound (\ref{DF}) becomes very stringent and implies that the 
$D_a$ are small. By the flatness condition we get then $F^i F_i \simeq 3\, m_{3/2}^2$, and the 
dynamical relation (\ref{DFF}) simplifies to 
\be
D_a \simeq \frac 12 \, m_{3/2}^{-2} \, q_{ai \barj} F^i F^{\barj} \,.
\label{DFFsmall}
\ee
Note that since $F_i \sim {\cal O}(m_{3/2})$, this expression has the form of a sum of terms that are 
${\cal O}(1)$, with ${\cal O}(q_{ai\barj})$ coefficients. But from the kinematical bound we know that the 
$D_a$ must actually be small, more precisely ${\cal O}(M_{ab})$. This implies that there must be a 
cancellation of the leading order contribution of the various terms: $q_{ai \barj} \, F^i F^{\barj} \simeq 0$. 
This shows that the vector multiplets can be very light only if the charged chiral multiplets contribute to 
supersymmetry breaking in a suitably aligned way. It also indirectly implies that, unless the charges 
$q_{a i \barj}$ become somehow degenerate, the vector fields can be much lighter than the gravitino 
only if the number of chiral multiplets is higher than the number of vector multiplets: $n > m$.

The fact that the relation between the $D_a$ and the $F_i$ simplifies in this case can be partly 
understood from the fact that in this limit supersymmetry is broken at a scale much higher than the 
scale of gauge symmetry breaking. The $D_a$ must then be small, as a consequence of the fact that 
supersymmetry breaking is dominated by the $F_i$ of the chiral multiplets. On the other hand, in order 
to approximately preserve gauge invariance, the vacuum expectations values of the latter cannot be 
arbitrary, but must rather satisfy $m$ approximate relations, which are identified with the conditions
$q_{ai \barj} \, F^i F^{\barj} \simeq 0$. 

\section{Constraints from flatness and stability}
\setcounter{equation}{0}

We want now to study the constraints that can be put on the theory by imposing the flatness condition,
guaranteeing the vanishing of the cosmological constant, and the stability condition, enforcing the 
positivity of the scalar mass matrix. In order to do so, it is useful to introduce the $2n$-dimensional 
vector of scalar fields $\phi^{\alpha} = \left(\phi^i \smallskip, \phi^{\bari}\right)$. Using this notation, 
the complete $2n$-dimensional mass matrix for the scalar fields can be written in terms of the 
blocks (\ref{mijbar}) and (\ref{mij}) as
\be
m^2_{\alpha \bar \beta} = \left(
\begin{matrix}
m^2_{i \barj} &  m^2_{ij} \smallskip \cr
m^2_{\bari \barj} & m^2_{\bari j}
\end{matrix}
\right) \,.
\label{VIJ}
\ee

In order to study the restrictions imposed by the requirement that the physical squared mass of the 
scalar fields are all positive, it is necessary to take appropriately into account the spontaneous 
breaking of gauge symmetries. In this process $m$ of the scalars, the would-be Goldstone 
bosons, are absorbed by the gauge fields and turned into their longitudinal components. In the 
unitary gauge, these modes are disentangled from the remaining physical scalar fields, and their 
$m$-dimensional mass matrix coincides with the mass matrix $M^2_{ab}$ of the gauge fields. 
Since this matrix is by construction positive definite, we do not need to worry about these directions 
for the stability of the physical scalar fields. What we have instead to impose is that the mass matrix 
$\tilde m^2\hspace{-4pt}{}_{\alpha \bar \beta}$ of the remaining $2n-m$ physical scalar fields is positive 
definite. This complicates the analysis with respect to the case where vector multiplets are 
absent, because the matrix $\tilde m^2\hspace{-4pt}{}_{\alpha \bar \beta}$ is defined by a complicated 
real projection of the Hermitian matrix $m^2\hspace{-4pt}{}_{\alpha \bar \beta}$ onto the subspace 
orthogonal to the would-be Goldstone directions $X^\alpha_a=(X^i_a,X^{\bar i}_a)$.

Fortunately, there exist a simple way to avoid the extra complications introduced by the gauge fixing
procedure. It relies on the observation that the would-be Goldstone modes correspond, before the 
gauge fixing, to flat directions of the unphysical mass matrix $m^2\hspace{-4pt}{}_{\alpha \bar \beta}$, 
and get their physical mass through their kinetic mixing with the gauge bosons 
\footnote{It is straightforward to verify that the Goldstone directions $X^\alpha_a$ correspond indeed 
to flat directions of the mass matrix $m^{2}_{\alpha \bar \beta} = \nabla_\alpha \nabla_{\bar \beta} V$. 
To see this note that, in the notation we are using in this section, the operator generating real gauge 
transformations is just $\delta = \lambda^a X_a^\alpha \nabla_\alpha$. The condition of gauge invariance 
of the potential $V$, namely $\delta V = 0$, implies thus that $X_a^\alpha \nabla_\alpha V = 0$. Since 
this relation is valid at any point, one can take another derivative and deduce the new condition 
$\nabla_{\bar \beta} X_a^\alpha \nabla_\alpha V + X_a^\alpha \nabla_{\bar \beta} \nabla_\alpha V = 0$. 
Now, evaluating this expression at a stationary point, where $\nabla_\alpha V = 0$, one finds 
$m^2_{\alpha \bar \beta} X_a^{\bar \beta} = 0$, implying in particular that 
$m^2_{\alpha \bar \beta} X_a^{\alpha} X_a^{\bar \beta} = 0$.}. 
This means that positivity of the physical mass matrix $\tilde m^2\hspace{-4pt}{}_{\alpha \bar \beta}$
implies semi-positivity of the unphysical mass matrix $m^2\hspace{-4pt}{}_{\alpha \bar \beta}$. 
This condition is much simpler to study, and we will therefore take it as our starting point.

To extract some useful information from the condition that the mass matrix (\ref{VIJ}) should be 
semi-positive definite, we can now use the same strategy as in \cite{grs1}. Instead of requiring 
that $m^2\hspace{-4pt}{}_{\alpha \bar \beta} \, Z^\alpha Z^{\bar \beta} \ge 0$ for any direction 
$Z^\alpha$, we can look only along some special directions, and see if we can obtain simple 
necessary conditions. Note first that combining the conditions coming from the two independent
directions $Z_1^\alpha = (z^i,z^{\bari})$ and $Z_2^\alpha = (iz^i,-iz^{\bari})$ defined by any complex
vector $z^i$, one deduces that $m^2\hspace{-4pt}{}_{i \barj} \, z^i z^{\barj} \ge 0$. This means 
that if $m^2_{\alpha \bar \beta}$ is semi-positive definite, then the principal block 
$m^2\hspace{-4pt}{}_{i \barj}$ must necessarily be semi-positive definite as well. 
Our strategy will then be, as in \cite{grs1}, to look for a suitable vector $z^i$ leading 
to a simple constraint on the parameters of the theory.

In the case of theories with both supersymmetry and gauge symmetries, there exist two types 
of special complex directions $z^i$ one could look at. The first is the direction $G^i$, which is associated 
with the Goldstino direction in the subspace of chiral multiplet fermions. The second is identified with the 
directions $X_a^i$, which are instead associated with the Goldstone directions in the space of chiral 
multiplet scalars. Consider first the direction $G^i$. Starting from (\ref{mijbar}) and using 
the fact that $G^i \nabla_i G_a = G_a$, the flatness condition (\ref{flatness}), and also the stationarity 
condition (\ref{stationarity}) to simplify terms involving $\nabla_i G_j$, one finds, after a long but 
straightforward computation:
\bea
\a\a m^2_{i \bar j} G^i G^{\barj} = e^G \Big[6 -R_{i \barj p \bar q} \, G^i G^{\barj} G^p G^{\bar q} \Big]
+\, \Big[\!-\! 2\, G^a G_a + h^{cd} h_{a c i} h_{b d \barj} \, G^i G^{\barj} G^a G^b \Big] \label{mGG} \\[0mm] 
\a\;\a \hspace{55pt} +\,e^{-G} \Big[M^2_{ab} G^a G^b \!-\! \frac 34\,Q_{abc} G^a G^b G^c 
\!-\! \frac 12 \Big(G^a G_a\Big)^2 \! \!+\! \frac 14 h_{ab}^{\;\;\;i} h_{cdi} G^a G^b G^c G^d \Big]\,. \nn 
\eea
The condition $m^2\hspace{-4pt}{}_{i \barj} \, G^i G^{\barj} \ge 0$ is then the generalization of the 
condition derived in \cite{grs1} for theories involving only chiral multiplets. We believe that also in this 
case this condition captures the most significant restrictions associated to stability, insofar it 
comes from the scalar and pseudoscalar partners of the Goldstino, which are generically expected 
to be the lightest and most dangerous modes.

Consider next the directions $X_a^i$. Using the fact that $X_a^i \nabla_i G_b = i/2\, M^2_{ab}$, as 
well as the condition (\ref{G1inv}) to simplify the terms involving $\nabla_i G_j$, one finds:
\bea
\a\a m^2_{i \bar j} X_a^i X_a^{\barj} = e^G \Big[M^2_{aa} - i\, q_{ai \barj} \Big(X_a^i G^{\barj} - X_a^{\barj} G^i \Big) 
+ q_{a i}^{\;\;\;k} q_{a \barj k} G^i G^{\barj} - R_{i \barj p \bar q} \, X_a^i X_a^{\barj} G^p G^{\bar q} \Big] \nn \\
\a\;\a \hspace{56pt} \,+\, \Big[ \frac 14 M^4_{aa} - \Big(q_{b i \barj} X_a^i X_a^{\barj} + \frac 12\, Q_{a b c} M^{2c}_a \Big) G^b 
- \frac 14 \, \Big(M^2_{aa} - 2\, G_a G_a\Big)G^b G_b \nn \\
\a\;\a \hspace{72pt} -\, M^2_{a b} G_a G^b + \frac 14 Q_{a b d} Q_{ac}^{\;\;\;d} G^b G^c 
+ \frac 12 Q_{abc} G_a G^b G^c \Big]\,.
\label{mXX}
\eea
The conditions $m^2\hspace{-4pt}{}_{i \barj} \, X_a^i X_a^{\barj} \ge 0$ represent in principle new 
additional constraints, which are absent in theories involving only chiral multiplets. We believe however
that these conditions do not capture any extremely relevant new information, as they involve
essentially the complex partners of the Goldstone scalars, which are not expected to be very 
dangerous  \footnote{This fact can be explicitly seen in the simplest case of models involving just one 
chiral and one vector field. In this case there are only two mass eigenvalues, which must therefore be 
in one-to-one correspondence with the two complex directions $X$ and $G'$. One of these eigenvalues does not vanish and 
is clearly associated to $G'$, whereas the other does vanish and is associated to $X$. It is then 
straightforward to verify that the non-zero eigenvalue is exactly captured by (\ref{mGG}), and that 
(\ref{mXX}) is just proportional to it.}.

We are now ready to use the conditions of flatness and stability at the stationary point to find some 
algebraic constraints applying to any supergravity theory, and which are simple enough to represent an 
interesting restriction on the functions $G$, $h_{ab}$ and $X_a$ that specify the theory. Recall that the flatness 
condition is given by eq.~(\ref{flatness}) and takes the form:
\be
- 3 + G^i G_i + \frac 12 \, e^{-G} G^a G_a = 0 \,.
\label{flat}
\ee
The necessary conditions for stability are more involved and correspond to requiring that the expressions 
(\ref{mGG}) and (\ref{mXX}) should be semi-positive. This leads to two kinds of algebraic conditions. 
Note however that (\ref{mGG}) depends only on $M^2_{ab}$ and not on $q_{ai\barj}$, and has a simple 
tensor structure, whereas (\ref{mXX}) depends both on $M^2_{ab}$ and $q_{ai\barj}$, and has a much 
more complicated tensor structure. As we already mentioned, we believe that only the former leads to 
a truly significant condition so we will therefore take as our necessary condition 
for meta-stability  the condition $m^2_{i \barj} \, G^i G^{\barj} \ge 0$, which implies:
\bea
\a\a R_{i \barj p \bar q} \, G^i G^{\barj} G^p G^{\bar q} \le 6
+\, e^{-G} \Big[\!-\! 2\, G^a G_a + h^{cd} h_{a c i} h_{b d \barj} \, G^i G^{\barj} G^a G^b \Big] \nn \\[0mm] 
\a\;\a \hspace{90pt} +\,e^{-2G} \Big[M^2_{ab}\, G^a G^b -\,\frac 34\,Q_{abc} \, G^a G^b G^c 
\label{stab}\\[0mm]
\a\;\a \hspace{130pt} -\frac 12 \Big(G^a G_a\Big)^2 \!+ \frac 14\, h_{ab}^{\;\;\;k} h_{cdk} \, G^a G^b G^c G^d  \Big]\,.\nn
\eea

The flatness condition (\ref{flat}) and the stability constraint (\ref{stab}) represent a 
strong restriction on the theory. They generalize the results derived in \cite{grs1} and \cite{grs2}
to the more general case where both chiral and vector multiplets participate significantly to supersymmetry
breaking. In the following, to work out more explicitly the implications of these constraints, 
we will mostly focus on the more tractable case where the gauge kinetic 
functions are constant and diagonal, so that $h_{ab} = g_a^{-2} \delta_{ab}$, 
$h_{a i} = 0$ and $h_{a i j} = 0$. In this situation, one has also necessarily $Q_{abc}=0$. 
The flatness and stability conditions take in this case  
the following simple form:
\bea
\left\{\hspace{-4pt}
\begin{array}{l}
\displaystyle{g_{i \barj} F^i F^{\barj} = 3\,m_{3/2}^2 - \frac 12\, h_{ab} \, D^a D^b \,,} \\[2mm]
\displaystyle{R_{i \barj p \bar q} \, F^i F^{\barj} F^p F^{\bar q} \le 6 \, m_{3/2}^4
+ \Big(M_{ab}^2 - 2\,m_{3/2}^2 h_{ab} \Big) D^a D^b}
- \frac 12\, h_{ab} h_{cd}\, D^a D^b D^c D^d \,.
\end{array}
\right.
\label{fscondFD}
\eea
Besides this, it is important to remember that the values of the $F_i$ and the $D_a$ are not 
completely independent, but rather related by eq.~(\ref{DFF}). In the simpler situation considered 
here, this relation implies, after using the flatness condition, that 
\footnote{Note that using the relations (\ref{DFFspec}) it is possible to rewrite the quadratic term 
$(M_{ab}^2 - 2\,m_{3/2}^2 h_{ab}) D^a D^b$ in (\ref{fscondFD}) as 
$2\, q_{a i\barj}\, D^a F^i F^{\barj} - 2\, g_{i \barj} h_{ab}\, F^i F^{\barj} D^a D^b$. One might then 
be tempted to further rewrite this as $2(Z_{i \barj ab} - g_{i \barj} h_{ab})\, F^i F^{\barj} D^a D^b$, 
in terms of the new quantity $Z_{i \barj ab} = D_{(a}^{-1} \nabla_i \nabla_j D_{b)}$, which does 
however not have any clear interpretation. This would allow us to rewrite the conditions (\ref{fscondFD}) 
as ${\cal G}_{\xi \bar \theta} \eta \raisebox{2.5pt}{$\scriptstyle \xi$} 
\eta  \raisebox{2.5pt}{$\scriptstyle \bar \theta$} = m_{3/2}^2$ and 
${\cal R}_{\xi \bar \theta \chi \bar \psi} \eta \raisebox{2.5pt}{$\scriptstyle \xi$} 
\eta  \raisebox{2.5pt}{$\scriptstyle \bar \theta$} \eta \raisebox{3.5pt}{$\scriptstyle \chi$} 
\eta  \raisebox{2.5pt}{$\scriptstyle \bar \psi$} \le 2/3 \, m_{3/2}^4$,
in terms of the Goldstino vector $\eta_\xi = (i F^i/\sqrt{3},D^a/\sqrt{6})$ and some generalized 
"metric" and "curvature" tensors. Unfortunately, it does not seem to be possible to write the 
conditions (\ref{DFFspec}) in any illuminating way in terms of $\eta_\xi$. One could then ignore 
(\ref{DFFspec}) and treat all the components of $\eta_\xi$ as independent variables, to define 
a weaker set of equations, in the same spirit as in \cite{grs2}, although we will not follow this direction 
here.}:
\be
q_{ai\barj} \, F^i F^{\barj} =  
\frac 12\,\Big[ M^2_{ab} + 4\, m_{3/2}^2 \, h_{ab} \Big]\, D^b - \frac 12\, h_{ab} h_{cd}\, D^b D^c D^d \,.
\label{DFFspec}
\ee

Recall finally that the auxiliary fields $F_i$ and $D_a$ are related to the complex directions $G^i$ 
and $X_a^i$ by the relations $F_i = - m_{3/2} G_i$ and $D_a = i X_a^i G_i$. This means that the 
$F_i$ are associated to the direction of supersymmetry breaking, whereas the $D_a$ are 
associated to the scalar products between this direction and the gauge symmetry breaking direction. 
Whenever the $D_a$ are small, the gauge symmetry breaking effects related to the vector multiplets
get weaker. More precisely, in the limit  $g_a D_a \to 0$ the eqs.~(\ref{fscondFD}) reduce to those 
emerging in the absence of vector multiplets. This is not at all obvious, since in this limit the $D$-terms 
still give a non-vanishing contribution to $m_{i \barj}^2$. As can be seen from (\ref{mijbar}) this contribution 
is however given simply by $g_a^2 X_{ai} X_{a\barj}$, and therefore vanishes along the direction $G^i$, 
as $g_a G^i X_{a i} = i g_a D_a$ and this vanishes in the limit $g_aD_a\to 0$. 
On the other hand, this limit implies also some constraints 
among the chiral multiplets. If the masses of the vectors are small ($g_a M_{ab} \to 0$), then the kinematical 
relation between $D_a$ and $F_i$ implies $g_a X_a^{i} F_i \to 0$, whereas if they are large 
($g_a M_{ab} \to \infty$), the dynamical relation (\ref{DFFspec}) between $D_a$ and $F_i$ implies 
instead $g_a q_{a i \barj} F^i F^{\barj} \to 0$. 

Finally we would like to briefly comment on the special case of models involving just one chiral 
multiplet and one vector multiplet. In this case the physical mass matrix has just $1$ entry and 
the condition for it to be positive is given exactly by (\ref{stab}). In this case, the symbol $\le$ in 
(\ref{stab}) and (\ref{fscondFD}) can therefore be replaced by $<$, and the condition is not only 
necessary but also sufficient.

\section{Analysis of the constraints}
\setcounter{equation}{0}

The flatness and stability conditions (\ref{fscondFD}) represent our main result regarding the constraints 
put on gauge invariant supergravity theories by the requirement that they should admit a phenomenologically
and cosmologically viable vacuum. Along this section we will address the 
problem of working out more concretely the implications of these constraints.

We assume for simplicity that the gauge group consists of only Abelian $U(1)$ factors. In such a case, 
it is always possible to choose the basis of vector fields so that the mass matrix of the gauge bosons is 
diagonal, $M^2_{ab} = M^2_a \delta_{ab}$. The gauge kinetic function is assumed to be diagonal as well, 
and therefore the gauge couplings can be written as $h_{ab} = g_a^{-2} \delta_{ab}$. We can then choose 
the special parametrization of scalar fields that corresponds to normal coordinates for the K\"ahler geometry, 
in which the K\"ahler metric becomes $\delta_{I \bar J}$. Similarly, we can rescale the vector fields to define flat 
indices also for vector indices, in such a way that the vector metric becomes just $\delta_{AB}$. This rescaling 
simply amounts to including a factor $g_a$ for each vector index $a$, and one has for instance 
$D_A = g_a D_a$ and $M^2_A = g_a^2 M_a^2$. In this way, no explicit dependence on $g_a$ is left in the 
formulae.

It is convenient to introduce new variables, obtained by suitably rescaling the auxiliary fields with the 
gravitino mass as follows:
\be\label{fd}
f^I = \frac {F^I}{\sqrt{3}\,m_{3/2}} \,,\;\hspace{1cm}\;
d^A = \frac {D^A}{\sqrt{6}\,m_{3/2}} \,.
\ee
Note that the flat capital indices of these new fields are raised and lowered with the diagonal metrics 
$\delta_{I \bar J}$ and $\delta_{AB}$. In addition, we also find convenient to introduce the following 
new quantities, defined by rescaling the Killing vectors and its derivatives by the vector masses:
\be
v_A^I = \frac {\sqrt{2} X_A^I}{M_A} \,,\;\hspace{1cm}\; 
T_{A I \bar J} = \frac {q_{A I \bar J}}{M_A} \,.
\ee
Finally, we shall introduce the following dimensionless parameters, measuring the hierarchies 
between the scales of gauge symmetry and supersymmetry breaking:
\be\label{ro}
\rho_{A} = \frac {M_{A}}{2\, m_{3/2}} \,.
\ee

In terms of the quantities we have just introduced, the conditions (\ref{fscondFD}) can be rewritten in 
the following form:
\bea
\left\{\hspace{-4pt}
\begin{array}{l}
\delta_{I \bar J}\, f^I f^{\bar J} = 1 - \summ_A d_A^2 \,, \\[2mm]
R_{I \bar J P \bar Q} \, f^I f^{\bar J} f^P f^{\bar Q} \le \displaystyle{\frac 23} 
+ \displaystyle{\frac 43} \, \summ_{A} \Big(2\, \rho_{A}^2 -1 \Big)\, d_A^2 
- 2\, \summ_{A,B} d_A^2 d_B^2 \,,
\end{array}
\right.
\label{condnew}
\eea
whereas the dynamical relation (\ref{DFF}), kinematical relation (\ref{DFkin}) and kinematical bound 
(\ref{DF}) read now:
\bea
\a\a d_A = \sqrt{\frac 32} \, \frac {\rho_A \,T_{A I \bar J}  \, f^I f^{\bar J}}
{\rho_{A}^2 - 1/2 + 3/2\, \delta_{I \bar J}\,f^I f^{\bar J}} \,, \label{DFFnew} \\
\a\a d_A = i \rho_A v_A^I f_I \,, \label{kinrelnew} \\[1.5mm]
\a\a |d_A| \le \rho_{A} \, \sqrt{\delta_{I \bar J}\, f^I f^{\bar J}} \,. \label{boundnew}
\eea
Notice that the bound (\ref{boundnew}) and the flatness condition (\ref{condnew}) imply, together, 
the following limits on the sizes of the auxiliary fields:
\be
\sqrt{\delta_{I \bar J}\, f^I f^{\bar J}} 
 \in  \Bigg[\frac 1{\sqrt{1+ \summ_B \rho_B^2}}, 1 \Bigg] \,, 
\;\hspace{1cm}\;
\sqrt{\summ_A d_A^2}  \in  \Bigg[0, \sqrt{\frac {\summ_B \rho_B^2 \raisebox{-4pt}{}}
{1+ \raisebox{11pt}{} \summ_B \rho_B^2} }\Bigg] \,.
\ee
The size of the $d_A$ auxiliary field is therefore limited by the kinematical bound for small $\rho_A$
and by the flatness condition for large $\rho_A$. 

To derive the implications of the constraints (\ref{condnew}), one should take into account the fact that  
$f^I$ and $d^A$ are not independent variables. There are then several possible strategies that one can 
follow. A first possibility is to use the the dynamical relation (\ref{DFFnew}) to write the $d_A$ in terms of 
the $f^I$. One can then consider the parameters $R_{I \bar J P \bar Q}$, $M_A$ and $T_{A I\bar J}$ as 
fixed and take as free variables the $f^I$. In this way the equations have a reasonably simple tensor 
structure, but they become of higher order in the variables. As a consequence, they can in principle be solved 
only in special limits on the values of the parameters, like for instance the limits of heavy and light 
vectors, where the relation (\ref{DFFnew}) simplifies and the effects induced by the $d_A$ represent only 
small corrections. A second possibility is to use the kinematical relation (\ref{kinrelnew}) to write the 
$d^A$ in term of the $f^I$. One can then consider the parameters $R_{I \bar J P \bar Q}$ and $X_A^I$ 
as fixed, and the quantities $f^I$ as free variables. The equations have then the same structure as for a 
theory with only chiral multiplets, with at most quartic terms, but their tensor structure is substantially more 
involved. Therefore in general the equations cannot be solved, and this type of analysis is efficient only 
in special cases, where $R_{I \bar J P \bar Q}$ and/or $X_A^I$ have special properties. A third possibility 
is to impose only the kinematical bound (\ref{boundnew}) to restrict the values of the $d^A$ in terms of 
the $f^I$. One can then consider the parameters $R_{I \bar J P \bar Q}$ and $M_A$ as fixed, and the 
quantities $f^I$ and $d^A$ as free variables subject only to a bound. By doing so, one clearly looses 
a substantial amount of information, and the resulting conditions will therefore be weaker than the 
original ones, although still necessary. This is similar in spirit to what was done in \cite{grs2} to study 
cases with only chiral multiplets but with a K\"ahler geometry leading to a complicated structure 
of the Riemann tensor $R_{I \bar J P \bar Q}$.

It is clear that switching from the dynamical relation (\ref{DFFnew}) to the kinematical relation (\ref{kinrelnew}) 
and finally to the kinematical bound (\ref{boundnew}) represents a gradual simplification of the formulae, which
is also accompanied by a loss of information. As a consequence, the three different types of strategies we 
just described, which are based on the use of these three different relations, will be tractable over an 
increasingly larger domain of parameters, but this will be accompanied by a gradual weakening of the 
implied constraints. In the following sections we will perform more explicitly these different analyses and 
see what kind of information can be obtained with each of them.

\section{Exploiting the dynamical relation between $d_A$ and $f_I$}
\setcounter{equation}{0}

As already discussed in section 4, whenever the masses $M_A$ of the vector multiplets are all much larger or much smaller than $m_{3/2}$, 
that is $\rho_A \gg 1$ or  $\rho_A \ll 1$, the situation simplifies. 
In those limits (and assuming that the quantities $T_{A I \bar J}$ remain finite) the auxiliary fields $d_A$ 
become small, and the dynamical relation (\ref{DFFnew}) simplifies. It becomes then convenient to use 
it and to apply the first strategy outlined in section 6.  

For the purpose of this section, it is useful to rewrite the conditions (\ref{condnew}) in a slightly different 
form, which exhibits in a better way the effect of the vector multiplets with respect to the situation involving 
only chiral multiplets. To do so, we define new rescaled variables in such a way to reabsorb the non-trivial 
right-hand side of the flatness condition, and to transfer all the non-trivial effect of the vector multiplets into 
the right-hand side of the stability condition. This can be done by introducing the following new quantities:
\be
z^I = \frac {f^I}{\sqrt{1 - \summ_A d_A^2}} \,.
\label{newvar}
\ee
Using these new variables $z^I$ instead of the $f^I$, the flatness and stability conditions
(\ref{condnew}) can be rewritten as:
\bea
\left\{\hspace{-4pt}
\begin{array}{l}
\displaystyle{\delta_{I \bar J}\, z^I z^{\bar J} = 1 \,,} \\[1mm]
\displaystyle{R_{I \bar J P \bar Q} \, z^I z^{\bar J} z^P z^{\bar Q} \le \frac 23 \, 
K(d_A^2,\rho_A^2) \,,}
\end{array}
\right.
\label{condnewnew}
\eea
where
\be
K(d_A^2,\rho_A^2) = 1 + 4\, \frac {\summ_A \rho_A^2 d_A^2 - \big(\summ_A d_A^2\big)^2 \raisebox{-5pt}{}}
{\big(1 - \summ_B d_B^2\big)^2} \,.
\label{Kdr}
\ee
Due to the non-linearity of the change of variable (\ref{newvar}), the dynamical relation (\ref{DFFnew}) 
becomes more complicated and does not allow us to explicitly express the variables $d_A$ in terms of 
the new $z_I$. Similarly, the kinematical relation (\ref{kinrelnew}) and the 
kinematical bound (\ref{boundnew}) also get modified. More explicitly, these three relations 
between auxiliary fields take now the following forms:
\bea
\a\a d_A\,\frac {1 + \rho_A^2 - 3/2\,\summ_B d_B^2 \raisebox{-5pt}{}}{1 - \summ_B d_B^2 \raisebox{11pt}{}} 
= \sqrt{\frac 32} \, \rho_A \,T_{A I \bar J}  \, z^I z^{\bar J} \,,
\label{DFFnewnew} \\[1mm]
\a\a \frac {d_A}{\sqrt{1 - \summ_B d_B^2}}= i \rho_A v_A^I z_I \,,
\label{kinrelnewnew} \\[-1mm]
\a\a \frac {|d_A|}{\sqrt{1 - \summ_B d_B^2}} \le \rho_A \,. 
\label{boundnewnew}
\eea
This formulation of the constraints becomes identical to the previous one for small $d_A$. It 
presents however some advantages. Note in particular that in the limit  $d_A \ll 1$ the new 
variables $z^I$ defined by (\ref{newvar}) approximately coincide with the $f^I$, and can be 
therefore efficiently used to study the leading order effects due to vector multiplets.
 
We will now show that for the cases we are going to study in this section, the conditions 
(\ref{condnewnew}) can be rewritten in the form:
\bea
\left\{\hspace{-4pt}
\begin{array}{l}
\displaystyle{\delta_{I \bar J}\, z^I z^{\bar J} = 1 \,,} \\[2mm]
\displaystyle{\tilde R_{I \bar J P \bar Q} \, z^I z^{\bar J} z^P z^{\bar Q} \le \frac 23 \,.}
\end{array}
\right.
\label{condsimp}
\eea
These equations have the same functional form as the ones found in the case with only chiral 
multiplets. However the relevant variables $z^I$ are rescaled with respect to the original variables 
$f_I$, and the coefficients $\tilde R_{I \bar J P \bar Q}$ are shifted with respect to the components 
of the Riemann tensor $R_{I \bar J P \bar Q}$.

\subsection{Heavy vector limit}

In the limit $\rho_A \gg 1$, the relation (\ref{DFFnewnew}) implies that the $d_A$ are small 
and given by:
\be
d_A \simeq \sqrt{\frac 32} \rho_{A}^{-1} \,T_{A I \bar J}\, z^I z^{\bar J} \,.
\label{DFFlargenew}
\ee
In the small $d_A$ limit the function $K$ in (\ref{Kdr}) can then be simplified by keeping only the leading 
term and evaluating it by using eq.~(\ref{DFFlargenew}). This results in a quartic dependence 
on the variables $z_i$. One can then rewrite the conditions in the form (\ref{condsimp}), with an 
effective curvature tensor given by:
\be
\tilde R_{I \bar J P \bar Q} \simeq R_{I \bar J P \bar Q} 
- 2 \, \summ_A \Big(T_{A I \bar J} \, T_{A P \bar Q} + T_{A I \bar Q} \, T_{A P \bar J} \Big) \,.
\label{Rnew}
\ee
In addition, note that since $d_A \simeq 0$, the kinematical relation (\ref{kinrelnewnew}) 
implies the $m$ constraints $v_A^I z_I \simeq 0$.

These formulae show that in the limit in which the vector multiplets are heavy, they give two kinds of effects.
On one hand, they induce a correction to the K\"ahler curvature for the chiral multiplets. On the other hand, 
they reduce the number of relevant variables $z^I$ coming from the chiral multiplets, as some of the directions 
are associated to the would-be Goldstone chiral multiplets that are absorbed by the vector multiplets. It is clear 
from the form of the new tensor (\ref{Rnew}) that the correction on the curvature induced by the vector multiplets 
is negative, so the net effect is (compared with the 
situation with just chiral fields) to help in fulfilling the bound (\ref{condsimp}). Note that the correction in (\ref{Rnew}) is not 
necessarily small and can be of ${\cal O}(1)$. Actually, even for $T_{A I \bar J} \sim {\cal O}(1)$, one has 
$z_I \sim {\cal O}(1)$ and $d_A \sim {\cal O}(\rho_A^{-1})$, so that the ratio $d_A/z_I \sim {\cal O}(\rho_A^{-1})$ is 
still small, as assumed. 

We can understand this result also from a slightly different point of view. We see from (\ref{Rnew}) that the 
correction induced by a heavy vector field to the K\"ahler curvature is negative and of order $M_A^{-2}$. 
This is by assumption small compared to $m_{3/2}^{-2}$, but it can still be large compared to $1$ (in Planck units). 
The original curvature $R$ of the chiral multiplet theory is instead of order $\Lambda^{-2}$, where $\Lambda$ 
is the scale of the most relevant higher-dimensional operator. One must certainly have $\Lambda \gg m_{3/2}$, 
and gravitational effects correspond to $\Lambda \sim 1$. The effects due to vector multiplets can thus really 
compete with the curvature effects due to chiral multiplets, or even dominate over them. In particular, when the 
original curvature effects are of gravitational origin, the vector multiplet effects become significant when 
$M_A \sim 1$. This is typically what happens for spontaneously broken anomalous $U(1)$ symmetries in 
string models.

As we explained in section 4, in the limit $\rho_A \gg 1$ it should be possible to integrate out the vector multiplets
and reinterpret their net effect through a correction to the K\"ahler potential of the chiral multiplets. It is interesting 
to verify that such a procedure indeed reproduces (\ref{Rnew}). To do so, we can use superfields and 
follow the same steps as in subsection 4.1. Recall in particular that the correction to the potential $G$ 
was shown to be equal to $\Delta G \simeq - \summ_A M_A^{-2} G_A^2$. Since in flat indices the Riemann 
tensor is given by the fourth derivative of the potential $G$, the corresponding correction to the curvature is:
\be
\Delta R_{I \bar J P \bar Q}  \simeq 
- \summ_A M_A^{-2} \nabla_I \nabla_{\bar J} \nabla_P \nabla_{\bar Q}  (G_A^2) \,.
\ee
Taking into account that $\nabla_I \nabla_J G_A = 0$ and $\nabla_I \nabla_{\bar J} G_A = - q_{A I \bar J}$, 
one recovers then the second term in the expression (\ref{Rnew}). Notice however that this simplified derivation of 
(\ref{Rnew}) is valid only under the assumption that the correction is small, whereas as already emphasized, 
(\ref{Rnew}) is actually valid also in more general situations where the correction can be sizable.

\subsection{Light vector limit}

In the limit $\rho_A \ll 1$, the bound (\ref{boundnewnew}) and the relation (\ref{DFFnewnew})
imply that the $d_A$ are small and given by
\be
d_A \simeq \sqrt{\frac 32}\, \rho_A\, T_{A I \bar J}\, z^I z^{\bar J} \,.
\label{DFFsmallnew}
\ee
Again, as in the previous subsection, we can simplify the function $K$ by keeping only the quartic term in 
$z_i$. This allows us to finally rewrite the conditions (\ref{condnewnew}) in the form (\ref{condsimp}), with 
an effective curvature tensor now given by:
\be
\tilde R_{I \bar J P \bar Q} \simeq R_{I \bar J P \bar Q} 
- 2\, \summ_A \rho_A^4 \Big(T_{A I \bar J} \, T_{A P \bar Q} + T_{A I \bar Q} \, T_{A P \bar J} \Big) \,.
\label{Rnewnew}
\ee
Therefore, the net effect of the vector multiplets is as before to reduce the K\"ahler curvature (with respect to 
the case with only chiral multiplets). In addition, the number of relevant directions in field space is 
again reduced, since the kinematical bound (\ref{boundnewnew}) together with the dynamical relation 
(\ref{DFFsmallnew}) imply the $m$ constraints $T_{A I \bar J} \, z^I z^{\bar J} \simeq 0$. 

\subsection{Arbitrary vector masses}

One may wonder whether it is possible to generalize the analyses done in subsections 7.1 and 7.2 to 
the case of arbitrary vector masses. In particular, the crucial question is whether the effect of vector 
multiplets could again be essentially encoded into a modification of the Riemann tensor. It is 
clear from the structure of the equations (\ref{condnewnew}) that in general this is not going to be 
the case. However, we will see that the original constraints imply some other weaker constraints,
which have indeed this form but contain less information except in the cases of 
large or small masses.

In the case of only one vector multiplet, the flatness condition guarantees that the numerator in 
(\ref{DFFnewnew}) never vanishes and stays always positive. Moreover, the kinematical bound 
(\ref{boundnewnew}) implies that it reaches its minimal value for $d=\rho/\sqrt{1 + \rho^2}$. 
Using this result, we can then get an upper bound for the quantity $|d|/(1-d^2)$ by evaluating 
the numerator at its minimum. In this way we find that:   
\be
\frac {|d|}{1-d^2} \le \sqrt{\frac 32} \, \frac {\rho \, \big(1 + \rho^2\big)}{1 + \rho^2/2+\rho^4} \, 
\big|T_{I \bar J} z^I z^{\bar J}\big|\,. 
\label{DFFsimpl}
\ee
We can now find an upper bound to the function $K$ by neglecting the negative term 
in (\ref{Kdr}) and using the bound (\ref{DFFsimpl}). We can finally substitute this upper 
bound for $K$ in the conditions (\ref{condnewnew}) and deduce a simpler but weaker set of the 
conditions. These have once again the form (\ref{condsimp}), where now:
\be
\tilde R_{I \bar J P \bar Q} = R_{I \bar J P \bar Q} 
- 2\, \epsilon(\rho) \,\Big(T_{I \bar J} \, T_{P \bar Q} + T_{I \bar Q} \, T_{P \bar J} \Big) \,,
\label{Rmore}
\ee
with a function $\epsilon(\rho)$ given by:
\be
\epsilon(\rho) = \frac {\rho^4 \, \big(1 + \rho^2\big)^2}{\big(1 + \rho^2/2+\rho^4\big)^2} \,.
\label{epsilon1}
\ee
This expression is valid for arbitrary $\rho$ and in the limits $\rho \gg 1$ and  $\rho \ll 1$ it correctly 
reproduces (\ref{Rnew}) and (\ref{Rnewnew}).

In the more general case of several vector multiplets, the situation is more complicated,
due to the fact that the numerator in the left-hand side of the dynamical relation (\ref{DFFnewnew}) 
can vanish \footnote{It is however clear that whenever this numerator becomes 
small, also the right-hand side of the equation must do so.}. For this reason it is no longer 
possible to find a simple bound of the type (\ref{DFFsimpl}) valid for all values of the 
parameters $\rho_A$. One can however still derive a useful bound in a large domain
of the space of values of the parameters $\rho_A$. To see which domain should be considered, 
note that the kinematical bound implies that:
\be\label{c}
1 + \rho_A^2 - \frac 32\,\summ_B d_B^2 \ge \frac 
{1 + \rho_A^2 - \summ_B \rho_B^2/2 + \rho_A^2 \, \summ_B \rho_B^2 \raisebox{-5pt}{}}
{1 + \summ_B \rho_B^2 \raisebox{11pt}{}} \,.
\ee
The right-hand side of (\ref{c}) is positive if 
\be\label{cc}
\rho_A^2 \ge \frac 14 \bigg(\!\!-\! 1 - 2\, \summ_{B\neq A} \rho_B^2 
+ \sqrt{4 \big(\summ_{B\neq A} \rho_B^2\big)^2 \! + 12\, \summ_{B\neq A} \rho_B^2 - 15} \, \bigg) \,.
\ee
The function in the right hand side of the inequality (\ref{cc}) vanishes when 
${\sum}_{B\neq A} \rho_B^2 = 2$. For 
lower values, ${\sum}_{B\neq A} \rho_B^2 <2$, the function is negative and the condition is thus 
always satisfied. On the other hand, for larger values, ${\sum}_{B\neq A} \rho_B^2  > 2$, the function 
is positive and there is thus a non-trivial restriction on the parameters. It actually turns out that in 
this case this function grows monotonically up to the maximal value of $1/2$, so if $\rho_A^2 > 1/2$ 
the condition (\ref{cc}) is automatically satisfied. This shows that the numerator in (\ref{c}) can 
vanish only in very special valleys of parameters space, where some of the $\rho_A$ are small 
and some other are large. On the contrary, if all of them are either not too small or no too large, 
then this problem does not appear. We can then restrict our analysis to the following two regions 
of parameter space:
\be
I_{>} = \Big\{\vec \rho \, \Big|\rho_A^2 \ge \frac 12 \,, \forall A\Big\} \,, \;\hspace{1cm}\;
I_{<} = \Big\{\vec \rho \, \Big|\rho_A^2 \le 2 \, \,, \forall A\Big\} \,.
\ee
In the domain $I_< \cup I_>$ one can then deduce the following 
simple upper bound for (\ref{DFFnewnew}):
\be
\frac {|d_A| \raisebox{-5pt}{}}{1- \summ_B d_B^2 \raisebox{10pt}{}} \le \sqrt{\frac 32} \, 
\frac {\rho_A \big(1 + \summ_B \rho_B^2\big)\raisebox{-5pt}{}} 
{1 + \rho_A^2 - \summ_B \rho_B^2/2 +  \rho_A^2\, \summ_B \rho_B^2 \raisebox{10pt}{}}
\big|T_{A I \bar J} z^I z^{\bar J}\big|\,.
\label{DFFsimplm}
\ee
This is the obvious generalization of the bound (\ref{DFFsimpl}) for the case of more than one 
vector multiplet. One can then proceed as before, and use this to derive an upper bound for 
the function $K$, which can then be substituted in the original conditions to deduce a simpler 
but weaker set of constraints. In this way one finds once again (\ref{condsimp}), with an effective 
curvature of the form
\be\label{ccc}
\tilde R_{I \bar J P \bar Q} = R_{I \bar J P \bar Q} 
- 2 \, \summ_A \epsilon(\rho_A) \,\Big(T_{A I \bar J} \, T_{A P \bar Q} + T_{A I \bar Q} \, T_{A P \bar J} \Big) \,,
\ee
where $\epsilon(\rho_A)$ is now given by
\be
\epsilon(\rho_A) = \frac {\rho_A^2 \big(1 + \summ_B \rho_B^2\big)^2 \raisebox{-5pt}{}} 
{\big(1 + \rho_A^2 - \summ_B \rho_B^2/2 +  \rho_A^2\, \summ_B \rho_B^2\big)^2 \raisebox{10pt}{}} \,. 
\label{epsilonm}
\ee
This expression correctly reproduces (\ref{Rnew}) and (\ref{Rnewnew}) in the limits $\rho_A \gg 1$ and  $\rho_A \ll 1$.
It also reproduces (\ref{epsilon1}) in the case of a single vector multiplet, since in that case $I_< \cup I_>$ coincides 
with the whole parameter space.

\subsection{Simple scalar geometries}

It was shown in \cite{grs1,grs2} that in the case of theories with only chiral multiplets the conditions of 
flatness and stability could be solved exactly for certain particular classes of scalar manifolds leading 
to a simple structure for the Riemann tensor. This is for instance the case when the scalar manifold 
factorizes in a product of one-dimensional scalar manifolds (factorizable scalar manifolds), 
or when it is a symmetric coset group manifold of the form $G/H$ (symmetric scalar manifolds). 
It is then illustrative and physically interesting to apply the results derived in this subsection to these simple examples. 

For factorizable manifolds, the metric is diagonal and therefore the structure of the Riemann tensor 
simplifies to $R_{I \bar J P \bar Q} = R_I \, \delta_{I \bar J P \bar Q}$. Assuming then also that the Killing 
potentials are separable, so that $T_{A I \bar J} = q_{A I}/M_A \, \delta_{I \bar J}$, the corrected 
effective curvature appearing in (\ref{ccc}) takes the following simple form:
\be
\tilde R_{I \bar J P \bar Q} = R_I \, \delta_{I \bar J P \bar Q} 
- 2\, \summ_A \epsilon(\rho_A) \frac {q_{A I} q_{A P}}{M_A^{2}} \,
\Big(\delta_{I \bar J} \, \delta_{P \bar Q} + \delta_{I \bar Q} \, \delta_{P \bar J} \Big) \,.
\ee
This is no longer separable, but the problem can however still be solved exactly. Indeed,
the flatness and stability conditions (\ref{condsimp}) reduce to
\bea
\left\{\hspace{-4pt}
\begin{array}{l}
\displaystyle{\summ_I |z^I|^2 \simeq 1 \,,} \\[2mm]
\displaystyle{\summ_{I,J} \tilde R_{IJ}\, |z^I|^2\, |z^J|^2
\le \frac 23 \,,}
\end{array}
\right.
\eea
where the $n$-dimensional matrix $\tilde R_{IJ}$ is given by
\be
\tilde R_{IJ} = R_I \, \delta_{IJ} - 4\, \summ_A \epsilon(\rho_A) \frac {q_{AI} \, q_{A J}}{M_A^2} \,.
\label{RIJ}
\ee
The values of the variables $z^I$ minimizing the stability condition subject to  
the flatness constraint are easily found to be $|z^I|^2 = \sum_J (\tilde R^{-1})_{IJ} / \sum_{R,S} (\tilde R^{-1})_{RS}$. Substituting these values into the stability bound, one finds then the following 
necessary condition:
\be
\summ_{I,J} (\tilde R^{-1})_{IJ} \ge \frac 32 \,.
\ee
Whenever the correction to the curvature induced by the vector multiplets is small, the inverse of the matrix
(\ref{RIJ}) is easily found and the above curvature constraint simplifies to:
\be
\summ_A \epsilon(\rho_A) \frac {(q_{AI} R_I^{-1})^2}{M_A^2}  \gsim \frac 14 \Big(\frac 32 - \summ_{I} R_I^{-1} \Big)\,.
\ee
It is then clear that in a case where $\sum_I R_I^{-1}$ is only slightly lower than $3/2$, 
the effect of vector multiplets can help in satisfying the bound.

For maximally symmetric scalar manifolds, the Riemann tensor is related to the metric and takes 
the form $R_{I \bar J P \bar Q} = R_{\rm all}/2 \, (\delta_{I \bar J} \delta_{P \bar Q} 
+\delta_{I \bar Q} \delta_{P \bar J})$, where $R_{\rm all}$ is an overall curvature scale. 
Assuming then as before that the Killing potentials are diagonal, $T_{A I \bar J} = q_{A I}/M_A \, \delta_{I \bar J}$, 
the corrected effective curvature takes the form:
\be
\tilde R_{I \bar J P \bar Q} = \Big(\frac {R_{\rm all}}{2} 
- 2\, \summ_A \epsilon(\rho_A)  \frac {q_{A I} \, q_{A P}}{M_A^2} \Big)\, 
\Big(\delta_{I \bar J} \, \delta_{P \bar Q} + \delta_{I \bar Q} \, \delta_{P \bar J} \Big) \,.
\ee
This is still of the maximally symmetric form, and the flatness and stability conditions (\ref{condsimp}) 
can then be written in the form:
\bea
\left\{\hspace{-4pt}
\begin{array}{l}
\displaystyle{\summ_I |z^I|^2 \simeq 1 \,,} \\[2mm]
\displaystyle{\summ_{I,J} \Delta \tilde R_{I \bar J} \, |z^I|^2\, |z^J|^2 \ge R_{\rm all} - \frac 23 \,,}
\end{array}
\right.
\eea
where the $n$-dimensional matrix $\Delta \tilde R_{I \bar J}$ is given by
\be
\Delta \tilde R_{I \bar J} = 4\, \summ_A \epsilon(\rho_A) \frac {q_{A I} \, q_{A J}}{M_A^2} \,.
\ee
The values of the $z^I$ that optimize the stability bound taking into account the flatness condition are again 
easily found, and read in this case $|z^I|^2 = \sum_J (\Delta \tilde R^{-1})_{IJ} / \sum_{R,S} (\Delta \tilde R^{-1})_{RS}$.
Substituting these values into the stability bound, one finds then the following necessary condition:
\be
\summ_{R,S} (\Delta \tilde R^{-1})_{RS} \le \Big(R_{\rm all} - \frac 23 \Big)^{-1} \,.
\ee
This means that whenever $R_{\rm all} \ge 2/3$, there need to be vector fields with sufficiently
low mass and large charges to avoid unstable modes. For instance, in the particular case with a 
single vector multiplet, the necessary condition is just $q_{\rm max}/M \ge 1/2 \sqrt{R_{\rm all} - 2/3}$.

\section{Exploiting the kinematical relation between $d_a$ and $f_I$}
\setcounter{equation}{0}

Whenever the Riemann tensor and the charge matrix have simple tensor structures, the kinematical 
relation (\ref{kinrelnew}) can become useful in the task of solving the constraints (\ref{condnew}). 
This is the case, for instance, when the scalar manifold is factorizable and the isometries that 
are gauged are each aligned along a single one-dimensional submanifold. In such a situation
the Riemann tensor is diagonal and has the form $R_{I \bar J P \bar Q} = R_I\,\delta_{I \bar J P \bar Q}$ 
and each Killing vector has a single non-vanishing component,
$X_A^I = X^I \, \delta_A^I$. In this situation it is convenient to use the second strategy outlined in 
section 6. In this case the number of chiral and vector multiplets are equal: $m=n$ and we can then 
identify the chiral and vector indices $I,J,\dots$ and $A,B,\dots$, keeping in mind that simpler cases 
with less vector multiplets can be described by simply setting some of the $X^I$ to 0  
\footnote{It should be emphasized that this simple example where the space factorizes in a number 
of sectors with one pair of chiral and vector multiplets each   is rather restrictive. Indeed, the requirement 
of gauge invariance of $G$ severely restricts the possible form of the superpotential $W$. In the case 
where both $K$ and $W$ are gauge invariant, $W$ can only be a constant, implying  that non-zero $f_I$ 
and $d_I$ can be induced only through a non-trivial $K$. A more general possibility is that $K$ and 
$W$ have gauge variations that compensate each other. This fixes however almost uniquely the form 
of $W$ in terms of the form of $K$, and one finds again a very similar situation.}. 

The crucial simplification in this case is that, due to the kinematical relation (\ref{DFkin}), each vector 
auxiliary field $d_I$ is proportional to the corresponding chiral auxiliary field $f_I$. For each 
one-dimensional subspace of the scalar manifold, the ratio of the two auxiliary fields is actually 
fixed by the mass of the corresponding vector field: $|d_I|/|f_I| = \rho_I$. The orientation of the 
Goldstino direction along the $n$ one-dimensional subspaces of the scalar manifold is instead 
arbitrary. It can be parametrized by $n$ variables $x_i$ related to the absolute sizes of $|f_i|$ 
and $|d_i|$ in each of these subspaces. A convenient choice is to define:
\be
x_I = \sqrt{|f_I|^2 + |d_I|^2} \,.
\ee
The flatness and stability conditions can then be rewritten in the following 
very simple form:
\bea
\left\{\hspace{-4pt}
\begin{array}{l}
\displaystyle{\summ_I x_I^2 = 1} \,, \\[2mm]
\displaystyle{\summ_{I,J} \tilde R_{IJ} \, x_I^2 \, x_J^2 + \frac 23\, \summ_I E_I \, x_I^2 \le \frac 23  \,,}
\end{array}
\right.
\label{condre}
\eea
where 
\be
\tilde R_{IJ} = \frac {R_I \, \delta_{IJ} + 2\, \rho_I^2  \rho_J^2}
{\big(1 + \rho_I^2 \big)\big(1 + \rho_J^2 \big)} \,, \;\hspace{1cm}\; 
E_I = \frac {2 \, \rho_I^2 \big(1 - 2\, \rho_I^2 \big) }{\big(1 + \rho_I^2 \big)} \,.
\ee
The problem can now be solved along the same lines as in \cite{grs1,grs2}. The constraint (\ref{condre}) 
can be interpreted as an upper bound on the function $f(x_I) = \sum_{I,J} R_{IJ}\, x_I^2\,x_J^2 + 
2/3 \sum_I E_I\,x_I^2$, where the real variables $x_I$ range from $0$ to $1$ and are 
subject to the constraint $\sum_K x_K^2 = 1$. In particular, the inequality (\ref{condre}) implies 
that $f_{\rm min}<2/3$, where $f_{\rm min}$ is the minimum value of $f(x_I)$ within the allowed range 
for the $x_I$. Finding $f_{\rm min}$ is a constrained minimization problem which can be solved in the 
standard way using Lagrangian multipliers. It is straightforward to show that the values of the variables 
at the minimum are given by 
\be
x_I^2 = \frac {1}{\summ_{P,Q} \tilde R_{PQ}^{-1}} 
\bigg[\summ_J \tilde R_{IJ}^{-1} - \frac 13\, \summ_{J,K,L} \tilde R_{IJ}^{-1} \tilde R_{KL}^{-1} \big(E_J - E_L \big) \bigg] \,.
\ee
The condition $f_{\rm min} < 2/3$ then implies that the constraint on the curvatures takes the form:
\be
\summ_{I,J} \tilde R_{IJ}^{-1} \big(1 - E_I \big) - \frac 1{6}\, \summ_{I,J,P,Q} \tilde R_{IJ}^{-1} \tilde R_{PQ}^{-1} 
\big(E_J - E_P \big) E_Q \ge \frac 32\,. 
\label{condit}
\ee
The expression (\ref{condit}) represents the main result of this section. It correctly reduces to the 
condition $\sum_I R_I^{-1} > 3/2$ in the limit where all the gauge couplings are switched off. 
The matrices $\tilde R_{IJ}$ are positive definite and reduce to $R_I \delta_{IJ}$ at zero coupling, 
whereas the parameters $E_I$ can be either positive or negative but vanish at zero coupling.

Using the expression of $\tilde R_{IJ}$, the inequality (\ref{condit}) can actually be rewritten more 
explicitly in terms of the inverse curvatures $R_I^{-1}$. The result takes the form:
\be
\summ_I \alpha_I\,  R_I^{-1} 
+ \summ_{I,J} \beta_{IJ}\, R_I^{-1} R_J^{-1}
+ \summ_{I,J,K} \gamma_{IJK}\, R_I^{-1} R_J^{-1} R_K^{-1}
\ge \frac 32 \,,
\label{condRRR}
\ee
where the coefficients $\alpha_I$, $\beta_{IJ}$ and $\gamma_{IJK}$ are positive and depend on the mass parameters 
$\rho_I$ as follows:
\bea
\a\a \alpha_I = 1+ 4\, \rho_I^6 \,, \nn \\[1mm]
\a\a \beta_{IJ} = \frac 83 \big(\rho_I^2 - \rho_J^2\big)^2 
\big(1 - \rho_I^2 - \rho_J^2 + \rho_I^4 + \rho_J^4 - 2 \rho_I^2 \rho_J^2 
+ 2\, \rho_I^4 \rho_J^2 + 2\, \rho_J^4 \rho_I^2 + \rho_I^4 \rho_J^4 \Big) \,, \nn \\
\a\a \gamma_{IJK} = \frac {16}3 \big(\rho_I^2 - \rho_J^2\big)^2 \big(\rho_I^2 - \rho_K^2\big)^2 
\big(\rho_J^2 - \rho_K^2\big)^2 \,.
\eea
Note that in the limit $\rho_I \rightarrow 0$ the constraint (\ref{condRRR}) reduces to the correct condition
$\sum_I R_I^{-1} \ge 3/2$ that was found in \cite{grs1,grs2} for the case of theories with only chiral multiplets. 
Actually, as all the coefficients $\alpha_{I}$, $\beta_{IJ}$ and $\gamma_{IJK}$ 
are positive and, in particular, $\alpha_I>1$ then the condition (\ref{condRRR}) 
is less stringent than the condition found in that case.  
Note also that for large $\rho_I$ the bound is trivially satisfied.

\subsection{One pair of chiral and vector multiplets}

In the simplest case of models involving $1$ chiral multiplet and $1$ vector multiplet, the 
situation is particularly simple. The flatness condition fixes $x = 1$. The auxiliary fields 
can then be parametrized as $|f| = \cos \delta$ and $|d| = \sin \delta$, where the angle $\delta$ is 
completely fixed by the parameter $\rho$ as $\rho = \tan \delta$. The condition (\ref{condRRR}) involves in this case only 
the first term so is linear in the inverse curvatures, and it implies the constraint:
\be
R \le \frac 23 \big(1 + 4 \tan^6 \! \delta \big) \,.
\ee
This shows that the stability condition can always be satisfied for sufficiently large values 
of $\delta$; more precisely, one needs $\delta \in [\delta_{\rm min}, \pi/2]$, where
\bea
\delta_{\rm min} = \left\{\hspace{-4pt}
\begin{array}{l}
\displaystyle{0 \,,\;\; R < \frac 23} \,, \\[2mm]
\displaystyle{{\rm arctan} \Big[\frac 38 \Big(R - \frac 23 \Big) \Big]^{1/6} \,,\;\; R > \frac 23} \,.
\end{array}
\right.
\eea
As expected, the bound $R < 2/3$ found in the limit of vanishing $\delta$ (in which the effect of the vector 
multiplet is negligible, and we recover the result found for just one chiral field), get corrected for 
non-vanishing $\delta$. Actually when the effect of the vector multiplet starts becoming important the bound 
gets milder and eventually trivializes for large $\delta$. Note however that the first correction appears 
only at sixth order in $\delta$.

\subsection{Two pairs of chiral and vector multiplets}

The next-to-simplest case is the case of models involving $2$ chiral multiplets and $2$ vector multiplets. 
The solution of the flatness condition can in this case be parametrized in terms of an arbitrary angle $\theta$ so that 
$x_1 = \cos \theta$ and $x_2 = \sin \theta$. The auxiliary fields can then be written as $f_1 = \cos \theta \cos \delta_1$, 
$d_1 = \cos \theta \sin \delta_1$, $f_2 = \sin \theta \cos \delta_2$, $d_2 = \sin \theta \sin \delta_2$,  where the 
angles $\delta_{1,2}$ are completely fixed by $\rho_{1,2} = \tan \delta_{1,2}$. The condition (\ref{condRRR}) 
involves in this case the first two terms, which are linear and quadratic in the curvatures, and implies a 
rather complicated constraint involving the curvatures $R_{1,2}$ and the angles $\delta_{1,2}$. From its 
structure, and the fact that the relevant coefficients $\alpha_I$ and $\beta_{IJ}$ are positive, it is however 
clear that this constraint is milder than the constraints that would arise for each pair of chiral and vector 
multiplets on its own. For example, in the particular case where the two sectors are identical one would find
the same constraint as for a single basic sector with a pair of chiral and vector multiplet, but with an effective 
curvature reduced by a factor of $2$, as is also the case when only chiral multiplets are present.

\subsection{Several pairs of chiral and vector multiplets}

In the more general case of models involving an arbitrary number of pairs of chiral and vector multiplets, 
the situation is even more complicated. Nevertheless it is interesting to point out that for situations where 
the parameters $\rho_I$ satisfy certain properties, it is possible to derive a condition that is linear, rather 
than cubic, in the inverse curvatures $R_I^{-1}$. More precisely, this is the case in the limit where all the 
$\rho_I$ are such that $\rho_I^2 \le 1/2$. In that case, the quantity $E_I$ is positive definite, so then it is 
possible to get a new (but weaker) condition by disregarding the term involving $E_I$ from the conditions 
(\ref{condre}). Following the same procedure that was used to derive eq.~(\ref{condRRR}), we get the 
necessary condition:
\be
\summ_I  \big(1+ \rho_I^2\big)^2\,  R_I^{-1} 
\ge \frac 32 \,.
\ee
This means that in this domain of parameters, namely $\rho_I^2 \le 1/2$, and the approximation considered here, 
the net effect of the vector multiplets is to effectively reduce the curvatures $R_I$ by a 
factor $(1+\rho_I^2)^{-2}$, that is, $\tilde R_I = \big(1 + \rho_I^2\big)^{-2} R_I $.

\section{Exploiting the kinematical bound between $d_a$ and $f_I$}
\setcounter{equation}{0}

In more general cases, for which the analyses of sections 7 and 8 cannot be applied, one may try
to work out the implications of the constraints (\ref{condnew}) by applying the third strategy outlined 
in section 6. This consists in considering both the $f^I$ and the $d^A$ as independent variables, 
constrained only by the kinematical bound (\ref{boundnew}). This approach is simple and can be 
worked out in full generality; however it clearly ignores a substantial amount of information 
concerning the actual dynamical and kinematical relations between the $f^I$ and the $d^A$. The 
resulting implications will therefore be weaker than the ones derived in the previous sections. 
Note in this respect that the kinematical bound represents a significant constraint only when $\rho_A \lsim 1$,
whereas it  becomes trivial when $\rho_A \gg 1$. We therefore expect that the condition resulting 
from this analysis will be stronger for small $\rho_A$ and will become weaker and weaker for increasing 
$\rho_A$.

For this analysis, it is convenient to use the new variables $z^I$ defined in (\ref{newvar}) instead 
of the $f_I$, and also similarly rescaled variables $\epsilon_A$ instead of the $d_A$. More precisely, 
we consider the following change of variables:
\be
z^I = \frac {f^I}{\sqrt{1 - \summ_A d_A^2}} \,, \;\hspace{1cm}\; 
\epsilon_A = \frac {d_A}{\sqrt{1 - \summ_B d_B^2}} \,.
\ee
Using these new variables, the flatness and stability conditions (\ref{condnewnew}) can be rewritten 
in the simple form:
\bea
\left\{\hspace{-4pt}
\begin{array}{l}
\displaystyle{\delta_{I \bar J}\, z^I z^{\bar J} = 1 \,,} \\[1mm]
\displaystyle{R_{I \bar J P \bar Q} \, z^I z^{\bar J} z^P z^{\bar Q} \le \frac 23 \, 
K(\epsilon_A^2, \rho_A^2) \,,}
\end{array}
\right.
\label{condbis}
\eea
where now
\be
K(\epsilon_A^2,\rho_A^2) = 1 + 4 \, \summ_A \rho_A^2 \, \epsilon_A^2
- 4 \, \summ_{A,B} \epsilon_A^2 \, \epsilon_B^2 + 4\, \summ_{A,B} \rho_A^2 \, \epsilon_A^2 \, \epsilon_B^2 \,,
\label{K}
\ee
and the kinematical bound (\ref{boundnew}) becomes simply:
\be
|\epsilon_A| \le \rho_A \,.
\label{boundbis}
\ee

It is now clear from (\ref{condbis}) that the most favorable situation is when the variables $z^I$ minimize the 
function $R_{I \bar J P \bar Q} \, z^I z^{\bar J} z^P z^{\bar Q}$ and the variables $\epsilon_A$ maximize the 
function $K(\epsilon_A^2,\rho_A^2)$. The maximization of $K(\epsilon_A^2,\rho_A^2)$ with respect to 
$\epsilon_A$ should be done taking into account the bound (\ref{boundbis}):
\be
K_{\rm best}(\rho_A^2) = {\rm max} \Big\{K(\epsilon_A^2,\rho_A^2) \,\Big|\, \epsilon_A^2 \le \rho_A^2 \Big\} \,.
\ee
The computation of this quantity is complicated by the fact that the function $K(\epsilon_A^2,\rho_A^2)$ 
depends on all the variables $\epsilon_A^2$, and that its maximum may lie inside region 
$\epsilon_A^2 < \rho_A^2$ or at the boundary $\epsilon_A^2 = \rho_A^2$, depending on the values of the 
parameters $\rho_A^2$. It is then a bit laborious, although straightforward, to characterize the constrained 
maximum of the function $K(\epsilon_A^2,\rho_A^2)$ over the full range of the parameters $\rho_A^2$. 
Once this maximal value $K_{\rm best}(\rho_A^2)$ has been found, one can substitute it in the 
stability condition to get the following necessary conditions:
\bea
\left\{\hspace{-4pt}
\begin{array}{l}
\displaystyle{\delta_{I \bar J}\, z^I z^{\bar J} = 1 \,,} \\[1mm]
\displaystyle{\tilde R_{I \bar J P \bar Q} \, z^I z^{\bar J} z^P z^{\bar Q} \le \frac 23 \,,}
\end{array}
\right.
\label{condtris}
\eea
where 
\be
\tilde R_{I \bar J P \bar Q} = K_{\rm best}(\rho_A^2)^{-1} R_{I \bar J P \bar Q} \,.
\label{NewR}
\ee
Again, from the structure of this constrain we see that the net effect of the vector multiplets is to reduce the effective 
curvature that is perceived by the chiral multiplets, in this case by the multiplicative factor $K_{\rm best}^{-1}$,
(which is clearly smaller than $1$). Unfortunately, it does not seem to be possible to find a closed  expression 
for $K_{\rm best}$ in general. We will thus first examine in detail the simplest models with $1$ and $2$ vector 
multiplets, and then discuss what can be said about the general case with $n$ vector multiplets.

\subsection{One vector field}

In the presence of one vector field, the function (\ref{K}) takes the form
\be
K(\epsilon^2,\rho^2) = 1 + 4 \, \rho^2 \, \epsilon^2 + 4 \, \big(\rho^2 - 1 \big)\, \epsilon^4 \,.
\label{K1}
\ee
The maximum of this function of $\epsilon^2$ within the region $[0,\rho^2]$ sits 
either at the stationary point $\rho^2/(2 - 2 \rho^2)$ or at the boundary
point $\rho^2$, depending on the value of $\rho^2$. More precisely,
if we define the two domains
\be
I_1 = \Big[0,\frac 12 \Big] \,, \; \hspace{1cm}\;
I_2 = \Big[\frac 12, +\infty\Big[ \,,
\ee
one finds:
\bea
{\epsilon^2}_{\rm best} = \left\{\hspace{-4pt}
\begin{array}{l}
\displaystyle{\frac 12 \, \frac {\rho^2}{1 - \rho^2} \,,\;\; \rho^2 \in I_1 \,,} \\[4mm]
\displaystyle{\rho^2 \,,\;\; \rho^2 \in I_2 \,.}
\end{array}
\right.
\eea
The corresponding maximal value of the function $K$ is then:
\bea
K_{\rm best} = \left\{\hspace{-4pt}
\begin{array}{l}
\displaystyle{\frac {1 - \rho^2 + \rho^4}{1 - \rho^2} \,,\;\; \rho^2 \in I_1} \,,\\[4mm]
\displaystyle{1 + 4\,\rho^6 \,,\;\; \rho^2 \in I_2 \,.}
\end{array}
\right.
\label{Kbest1}
\eea
Note that when a single chiral multiplet is present, the kinematical relation fixes 
$\epsilon^2 = \rho^2$, and one has to take the second branch of the expression 
(\ref{Kbest1}) for any value of $\rho$. On the other hand, when two or more chiral 
multiplets are present, one can in principle have $\epsilon^2 < \rho^2$, and one 
needs to consider the first branch of (\ref{Kbest1}) for small values of $\rho$.

\subsection{Two vector fields}

In the presence of two vector fields, the function (\ref{K}) takes the form
\bea
\a\a K(\epsilon_{1,2}^2) = 1 + 4 \, \big(\rho_1^2 \, \epsilon_1^2 + \rho_2^2 \, \epsilon_2^2\big) 
+ 4 \, \big(\rho_1^2 - 1 \big)\, \epsilon_1^4 + 4 \, \big(\rho_2^2 - 1 \big)\, \epsilon_2^4 \nn \\[1mm]
\a\;\a \hspace{48pt} +\, 4\, \big(\rho_1^2 + \rho_2^2 - 2 \big) \, \epsilon_1^2 \, \epsilon_2^2 \,.\label{uno}
\eea
The problem is symmetric, and we can thus assume without loss of generality that $\rho_1^2 \le \rho_2^2$.
The maximum of the function (\ref{uno}) with respect to ($\epsilon^2_1$,$\epsilon^2_2$) within the 
region $[0,\rho_1^2] \times [0,\rho_2^2]$ sits at the points $(\rho_1^2/(2 - 2\rho_1^2), 0)$ and 
$(\rho_1^2, \rho_2^2 + \rho_1^2 (\rho_1^2 + \rho_2^2 - 2)/(2 - 2\rho_2^2))$, or at the boundary points 
$(\rho_1^2,0)$ and $(\rho_1^2,\rho_2^2)$. More precisely, the relevant domains turn out to be:
\bea
\a\a I_A = \bigg\{ (\rho^2_1,\rho_2^2) \,\Big|\, 
\rho_1^2 \in \Big[0, \frac 12\Big] \,,\, 
\rho_2^2 \in \Big[0, \rho_1^2\Big]  \bigg\} \,, \\
\a\a I_B = \bigg\{ (\rho^2_1,\rho_2^2) \,\Big|\, 
\rho_1^2 \in \Big[\frac 12,2\Big] \,,\, 
\rho_2^2 \in \Big[0, \frac {2 - \rho_1^2}{1+ \rho_1^2} \rho_1^2 \Big]  \bigg\} \,, \\
\a\a I_C = \bigg\{ (\rho^2_1,\rho_2^2) \,\Big|\, 
\rho_1^2 \in \Big[\frac 12,\frac 2{\sqrt{7}}\Big] \,,\, 
\rho_2^2 \in \Big[\frac {2 - \rho_1^2}{1+ \rho_1^2} \rho_1^2, \rho_1^2 \Big]  \bigg\} \,, \\
\a\a I_D = \bigg\{ (\rho^2_1,\rho_2^2) \,\Big|\, 
\rho_1^2 \in \Big[\frac 2{\sqrt{7}},2\Big] \,,\, 
\rho_2^2 \in \Big[\frac {2 - \rho_1^2}{1+ \rho_1^2} \rho_1^2, \frac {1 - \rho_1^2 + \sqrt{1 + 14 \rho_1^2 - 7 \rho_1^4}}4 \Big] \bigg\} \,, \\
\a\a I_E = \bigg\{ (\rho^2_1,\rho_2^2) \,\Big|\, 
\rho_1^2 \in \Big[\frac 2{\sqrt{7}},2\Big] \,,\, 
\rho_2^2 \in \Big[\frac {1 - \rho_1^2 + \sqrt{1 + 14 \rho_1^2 - 7 \rho_1^4}}4, + \infty \Big[ \bigg\} \,, \\
\a\a I_F = \bigg\{ (\rho^2_1,\rho_2^2) \,\Big|\, 
\rho_1^2 \in \Big[2,+\infty\Big[ \,,\, 
\rho_2^2 \in \Big[0, + \infty \Big[ \bigg\} \,,
\eea
and the constrained maximum is located at the following points:
\bea
\big({\epsilon^2_1}, {\epsilon^2_2}\big)_{\rm best}  = \left\{\hspace{-4pt}
\begin{array}{l}
\displaystyle{\Big(\frac 12 \, \frac {\rho^2_1}{1 - \rho^2_1}, 0 \Big) \,,\;\; 
\big(\rho^2_1,\rho_2^2\big) \in I_A} \,, \\[4mm]
\displaystyle{\big(\rho^2_1, 0 \big) \,,\;\; 
\big(\rho^2_1,\rho_2^2\big) \in I_B} \,, \\[3mm]
\displaystyle{\Big(\rho_1^2, \frac 12\, \frac {\rho_2^2 + \rho_1^2 (\rho_1^2 + \rho_2^2 - 2)}{1 - \rho_2^2} \Big) 
\,,\;\; \big(\rho^2_1,\rho_2^2\big) \in I_C \cup I_D} \,, \\[4mm]
\displaystyle{\big(\rho_1^2, \rho_2^2 \big) 
\,,\;\; \big(\rho^2_1,\rho_2^2\big) \in I_E \cup I_F} \,. \\
\end{array}
\right.
\eea
The corresponding maximal value of the function $K(\epsilon_{1,2}^2)$ is:
\bea
K_{\rm best}  = \left\{\hspace{-4pt}
\begin{array}{l}
\displaystyle{\frac {1 - \rho^2_1 + \rho^4_1}{1 - \rho^2_1} \,,\;\; 
\big(\rho^2_1,\rho_2^2\big) \in I_A} \,, \\[4mm]
\displaystyle{1 + 4\, \rho_1^6 \,,\;\; 
\big(\rho^2_1,\rho_2^2\big) \in I_B} \,, \\[3mm]
\displaystyle{\frac {1 - \rho^2_2 + \rho^4_2 + \rho_1^2 \big(\rho_1^2 + 2\big)\rho_2^2\big(\rho_2^2-2\big) 
- 2 \rho_1^6 \rho_2^2 + \rho_1^8}{1 - \rho^2_2}
\,,\;\; \big(\rho^2_1,\rho_2^2\big) \in I_C \cup I_D}  \,, \hspace{-20pt} \\[4mm]
\displaystyle{1 + 4 \big(\rho_1^6 + \rho_2^6 \big) 
+ 4 \rho_1^2 \rho_2^2 \big(\rho_1^2 + \rho_2^2 - 2 \big)
\,,\;\; \big(\rho^2_1,\rho_2^2\big) \in I_E \cup I_F} \,. \\
\end{array}
\right.
\label{Kbest2}
\eea

\subsection{Arbitrary number of vector fields}

For a larger number of vector multiplets, the maximization problem becomes more and more
complex, and it does not seem to be possible to find the general solution. Nevertheless, it turns 
out to be possible to make some quantitative analysis also in this more general case, by either
making some assumptions on the parameters $\rho_A$ or by weakening the conditions.

In the particular case where all the parameters $\rho_A$ are such that $\rho_A > 1$ 
 it is clear from the form of the expression (\ref{K}) that the constrained maximum of 
the function $K(\epsilon_A^2,\rho_A^2)$ is located at the corner of the parameter space where 
all the $\epsilon_a^2$ are maximal, and therefore ${\epsilon_A^2}_{\, \rm best} = \rho_A^2$. 
In terms of the variables $d_A$, this means that the preferred situation is the one in which 
$d_A^2 = \rho_A^2/(1 + \sum_B \rho_B^2)$. Note that this value is in general less than the 
maximal value that each $d_A^2$ is individually allowed to take by the kinematical bound, 
but their sum is instead equal to the maximal value $\sum_A d_A^2=\sum_A \rho_A^2/(1 + 
\sum_B \rho_B^2)$ it is allowed to take. In other words, the optimal situation is realized 
in one particular direction in the space of the $d_A$ such that $\sum_A d_A^2$ is maximal. 
The corresponding value of the function $K$ is in this case:
\be
K_{\rm best} = 1 + 4 \, \summ_A \rho_A^4 - 4\, \summ_{A,B} \rho_A^2\, \rho_B^2 
+ 4\, \summ_{A,B} \rho_A^4 \, \rho_B^2 \,.
\label{Kunconstr}
\ee
Note that this expression correctly reproduces that last branches of the complete 
results (\ref{Kbest1}) and (\ref{Kbest2}) for $1$ and $2$ vector multiplets.

For completely generic values of the parameters $\rho_A$, it is possible to extract a weaker 
information from the constraints (\ref{condbis}) by noticing that 
\be
K(\epsilon_A^2,\rho_A^2) \le 1 + 4\, \rho^2 \, \epsilon^2 + 4\, (\rho^2 -1)\, \epsilon^4 \,,
\label{Kupper}
\ee
with:
\be
\label{sumvar}
\epsilon^2 = \summ_A \epsilon_A^2\,,\;\hspace{1cm}\;\rho^2 = \summ_A \rho_A^2\,.
\ee
One can then use the upper bound (\ref{Kupper}) instead of the original expression 
(\ref{K}) in the constraints. Doing so, one clearly looses any distinction among the various 
different multiplets, and one obtains instead a condition where the effect of the multiplets is somehow
averaged. In fact, the expression (\ref{Kupper}) has the same form as the expression 
(\ref{K1}) that is valid in the case of a single vector multiplet, but with $\epsilon^2$ and 
$\rho^2$ given by eqs.~(\ref{sumvar}). The constrained maximum of (\ref{Kupper}) within
the range $\epsilon^2 \le \rho^2$ is located at
\bea
{\epsilon^2}_{\rm best} = \left\{\hspace{-4pt}
\begin{array}{l}
\displaystyle{\frac 12 \, \frac {\rho^2}{1 - \rho^2} \,,\;\; \rho^2 < \frac 12} \,, \\[4mm]
\displaystyle{\rho^2 \,,\;\; \rho^2 > \frac 12 \,.} 
\end{array}
\right.
\eea
In terms of the original variables, this means the optimal situation is the one in which  
all the $d_A$ are such that the quantity $\sum_A d_A^2$ takes the 
maximal value $\sum_A \rho_A^2/(1 + \sum_B \rho_B^2)$ that it is allowed by 
the kinematical bound. As we already mentioned, there is a whole 
$m$-sphere of such directions, and no specific direction is singled out as optimal. 
This is due to the fact that in the present analysis any information distinguishing 
the different vector multiplets has been disregarded from the beginning.
For the function $K$ one finds
\bea
K_{\rm best} \le \left\{\hspace{-4pt}
\begin{array}{l}
\displaystyle{\frac {1 - \rho^2 + \rho^4}{1 - \rho^2} \,,\;\; \rho^2 < \frac 12} \,, \\[4mm]
\displaystyle{1 + 4\,\rho^6 \,,\;\; \rho^2 > \frac 12 \,.}
\end{array}
\right.
\eea

A last way to derive a weaker constraint that is valid for arbitrary values of the parameters 
$\rho_A$ is to neglect the negative term in $K$. One is then left with a monotonically 
growing function of the variables $\epsilon_A$, which obviously takes its maximal value 
at the boundary $\epsilon_A = \rho_A$ of the allowed domain. This leads to:
\be
K_{\rm best} = 1 + 4 \, \summ_A \rho_A^4 + 4\, \summ_{A,B} \rho_A^4 \, \rho_B^2  \,.
\ee
This condition is completely general, and still keeps a distinction between the different vector
multiplets. However, it is generically much weaker that the conditions we got in the 
previous analysis.

\subsection{Simple scalar geometries}

It is worth discussing more precisely the minimization of the term involving the effective 
Riemann tensor in the stability condition in the particular cases of separable or symmetric manifolds  
where, as already mentioned in section 7, the situation simplifies \cite{grs1,grs2}. 

For factorizable manifolds, the Riemann tensor can be written in terms of the $n$ curvature
scalars $R_I$ of the one-dimensional factors and is just given by 
$R_{I \bar J P \bar Q} = R_I \, \delta_{I \bar J P \bar Q}$. The effective curvature that is relevant 
for the constraints (\ref{condtris}) has then the same factorized form, and is given by
$R_{I \bar J P \bar Q} = \tilde R_I \, \delta_{I \bar J P \bar Q}$,
where 
\be 
\tilde R_I = K_{\rm best} (\rho_A^2)^{-1} R_I \,.
\ee
The values of the variables $z^I$ optimizing (\ref{condtris}) are given by 
$|z^I|^2 = \tilde R_I^{-1} /\sum_J \tilde R_J^{-1}$. In terms of the original variables, 
this means that the $f^I$ align along the direction with highest total effective inverse curvature. 
One finds then the following necessary condition on the effective inverse curvature scalars:
\be
\summ_I \tilde R_I^{-1} \ge \frac 32 \,.
\ee

For maximally symmetric manifolds, the Riemann tensor can be written in terms of a 
single curvature scale $R_{\rm all}$ as $R_{I \bar J P \bar Q} = R_{\rm all}/2 \, 
(\delta_{I \bar J} \delta_{P \bar Q} + \delta_{I \bar Q} \delta_{P \bar J})$. 
The effective curvature appearing in the flatness and stability conditions (\ref{condtris}) 
remains again of the same form, and reads $\tilde R_{I \bar J P \bar Q} = \tilde R_{\rm all}/2 \, 
(\delta_{I \bar J} \delta_{P \bar Q} + \delta_{I \bar Q} \delta_{P \bar J})$, where:
\be
\tilde R_{\rm all} = K_{\rm best} (\rho_A^2)^{-1} R_{\rm all} \,.
\ee
In this case, no particular direction in the variables $z^I$, or equivalently in the $f^I$, 
is singled out as optimal, due to the fact that the space is maximally 
symmetric and all the directions are thus equivalent. One finds however the necessary 
condition:
\be
\tilde R_{\rm all}^{-1} \ge \frac 32 \,.
\ee

\section{String inspired examples}
\setcounter{equation}{0}

In this section we will apply our results to the typical situations arising for the moduli sector of string models.
In order to get a better idea of the usefulness of the conditions (\ref{condnew}) for finding phenomenologically 
viable vacua in the presence of $D$-terms, we will study in more detail the form that these 
conditions take in the simplest case involving $1$ chiral and $1$ vector multiplet, and in the next-to-simplest
case of $2$ chiral multiplets with a factorized geometry and $1$ vector multiplet. We will also briefly 
comment on how the idea of uplifting introduced in \cite{KKLT}  fits into our study. 

\subsection{One chiral superfield and one isometry}

The simplest possible situation arises in a low energy effective theory 
with one chiral superfield and one isometry, which is 
gauged with one vector multiplet. For moduli fields of string models, the prototype of K\"ahler potential 
describing such a situation is of the form:
\be\label{k1}
K = - n\, {\rm log} \big(\Phi + \Phi^\dagger \big)\,. 
\ee
This is a constant curvature manifold with $R = 2/n$. It has a global symmetry associated to 
the Killing vector $X = i\, \xi$, which can be gauged as long as the superpotential is also 
gauge invariant. 

In this case, the flatness condition can be solved by introducing an angle $\delta$ and parametrizing 
the auxiliary fields defined in (\ref{fd}) as $f = \cos \delta$ and $d = \sin \delta$. The angle $\delta$ is, in 
this simple case, fixed by the ratio between the vector and the gravitino masses, the parameter 
$\rho$ defined in (\ref{ro}):
\be
{\rm tan}\,\delta = \rho \,. 
\ee
In this simple situation, the stability condition is actually necessary and sufficient for the only 
non-trivial mass eigenvalue to be positive. We can thus change the $\le$ sign to a $<$ sign, 
and write this condition as
\be
R\, <\, \frac23 \big(1+4\, \rho^6\big)\,.
\label{c1}
\ee
From this expression, it is clear that it is always possible to satisfy the stability condition with 
a large enough value of $\rho$. More precisely, recalling that $R=2/n$, we get that
\be
n > \frac {3}{1 + 4\, \rho^6} \,.
\label{c2}
\ee
Note in particular that eq.~(\ref{c2}) implies that whenever $n$ is substantially less than 
$3$, which is the critical value for stability in the absence of gauging, the contribution to 
supersymmetry breaking coming from the $D$ auxiliary field must be comparable to the 
one coming from the $F$ auxiliary field. This would be for instance the case for an effective
theory based on the dilaton modulus, for which $n=1$. On the other hand, if $n$ is close 
or equal to $3$, like in the case of an effective theory based on the volume modulus, 
the effect of $D$ auxiliary fields can be still relevant even if supersymmetry breaking is 
dominated by the $F$ auxiliary field.

It is important to note that in this case gauge invariance of $G$ severely restricts the form of the 
superpotential, which can only be of the form $W = \alpha \, e^{- \beta \, \Phi}$. With 
these forms of $K$ and $W$, it is easy to see that a satisfactory extremum can actually 
exist only if $n < 3$, as in the absence of the $D$-term we get a negative definite ($n<3$) 
or positive definite ($n>3$) scalar potential and we need the negative one to compete with 
the positive definite contribution due to the $D$-terms in order to stabilize the field. 
This model was studied in detail in \cite{vz1}
\footnote{In the more general case of a non-constant gauge kinetic function, which was 
considered in \cite{vz1}, the condition (\ref{c1}) gets modified. In terms of the parameter
$\kappa = g'/(g \sqrt{K''})$, one finds:
\be
R < \frac23 \big(1+4\,\rho^6\big) + 4 \,\sqrt{3}\,\kappa\, \rho^4 \sqrt{1 + \rho^2}
+ 4 \,\kappa^2\, \rho^2 \big(2 + \rho^2 \big) \,. 
\label{c3}
\ee
Again, this condition can always be satisfied for appropriate values of $\delta$.}. Of course, 
as was already noted in \cite{vz1}, the concept of uplifting a supersymmetric minimum using $D$-terms 
does not apply in this simple case as if we consider only one chiral field 
and one vector field the $D$-term is proportional to the only available $F$-term.

\subsection{Two chiral superfields and one isometry}

Another interesting and reasonably simple case is that of a low energy effective theory with 
two chiral superfields with a factorizable 
geometry and one isometry, which is gauged with one vector multiplet. For moduli fields 
in string models, one typically has a K\"ahler potential of the form 
\be\label{k2}
K = -n_1\, {\rm log}\, (\Phi_1+\Phi^\dagger_1)\,-n_2 \,{\rm log}\,(\Phi_2+\Phi^\dagger_2)\,.
\ee
The scalar manifold defined by this K\"ahler potential consists of two one-dimensional 
subspaces with constant scalar curvatures $R_1= 2/n_1$ and $R_2 = 2/n_2$,  and has 
actually two independent global symmetries, under which the chiral multiplets independently 
shift by an imaginary constant. We will consider the case in which we gauge only one linear 
combination of the isometries, defined by $X = (i\,\xi_1,i\,\xi_2)$. The most general superpotential 
allowed by gauge invariance of $G$ is then of the form 
$W = e^{-\beta (\xi_1\Phi_1+ \xi_2 \Phi_2)} W(\xi_2\Phi_1-\xi_1\Phi_2)$. 

In this case the solution of the flatness condition can be parametrized by two angles $\delta$
and $\theta$, with $f_1 = {\rm cos}\,\theta \cos \delta$, $f_2 = {\rm sin}\,\theta \cos \delta$ and 
$d = \sin \delta$.  It is also useful to introduce the following two angles, defining the orientation 
of the Killing vector $X = (X^1,X^2)$ and the inverse curvature vector $R^{-1} = (R_1^{-1},R_2^{-1})$:
\be
{\rm tan}\,x = \frac {\xi_2}{\xi_1} \,, \;\hspace{1cm}\;
{\rm tan}\,r = \frac {n_2}{n_1} \,.
\ee 
From the definitions of the vector mass, the gravitino mass, and the parameter $\rho$, 
one finds in this case the following relation between the angles $\delta$ and $\theta$:
\be
\frac{{\rm tan}\,\delta}{{\rm cos}\,(\theta-x)} = \rho \,.
\ee
The stability condition can then be written in the following form:
\bea
\label{3}
\a\a \Big(R_{1} \,{\rm cos}^4 \theta + R_{2} \,{\rm sin}^4 \theta\Big) 
+ \frac 23 \big(\rho^4 \hspace{-1pt} - \hspace{-1pt} \rho^6\big) \! 
\left(2\,{\rm cos}^2(\theta \hspace{-1pt} - \hspace{-1pt} x) 
- \frac {1}{1 \hspace{-1pt} - \hspace{-1pt} \rho^2} \right)^2 \! 
\le \frac 23 \,\frac{1 \hspace{-1pt} - \hspace{-1pt} \rho^2 \hspace{-1pt} + \hspace{-1pt} \rho^4}
{1 \hspace{-1pt} - \hspace{-1pt} \rho^2}\,.
\eea
In general, this condition cannot be solved exactly, since it is quartic. It is however easy 
to see that the first term in the left-hand side of (\ref{3}) takes its minimal value when 
${\rm cos}^2\theta = R_2/(R_1+R_2)$, implying $\theta=r$, whereas the second term gets 
minimized when ${\rm cos}^2(\theta-x) = 1/[2(1-\rho^2)]$ if $\rho \le 1/2$ and 
when $\theta=x$ if instead $\rho> 1/2$. One can then derive a lower bound for the left-hand 
side by minimizing the two terms separately, and derive in this way a simple necessary  
condition. Recalling that $R_1=2/n_1$ and $R_2=2/n_2$, one deduces in this way that:
\bea
\label{ar}
n_1+n_2 \ge
3 \left\{\!\!\!
\begin{array}{lll}
\displaystyle{\frac {1-\rho^2}{1-\rho^2+\rho^4}} \,,
&\mbox{if}& \rho \le 1/2\,, \hspace{-30pt} \bigskip\ \\[0mm]
\displaystyle{\frac {1}{1+4\,\rho^6}}\,, &\mbox{if}& \rho > 1/2\,.  \hspace{-30pt} \smallskip\ \\
\end{array}\right.
\eea
It is clear from this condition that for large enough $\rho$ one can always satisfy the bound, 
the minimal value needed for $\rho$ being determined by $n_1$ and $n_2$. Note also that the 
expression (\ref{ar}) reproduces the same bounds derived in section 9, as is clear from 
eq.~(\ref{Kbest1}).

An interesting particular case occurs when one of the shifts vanishes, say $\xi_1 = 0$. In such 
a situation, one finds $x = \pi/2$, and the condition (\ref{3}) can be solved exactly. Actually it turns out that for values of $\rho$ such that $3\,(n_1+n_2)+4 \,n_1n_2\,\rho^4\,(1-\rho^2) \le 0$, the bound can always be satisfied 
for some range of values of the angle $\theta$. On the other hand, for values of $\rho$ such that 
$3\,(n_1+n_2)+4\, n_1n_2\,\rho^4\,(1-\rho^2) > 0$ the following condition must be 
satisfied:
\be
\label{C2}
n_1 + n_2 \ge 3 - 4 \, n_2 \, \rho^6 - \frac 43 \, n_1 n_2 \, \rho^4 \big(1 - \rho^2 + \rho^4 \big) \,.
\ee
Note also that in such a model, it is possible to stabilize the field $\Phi_1$ in a supersymmetric 
way. This would correspond to $\theta=\pi/2$, and the situation would then become identical 
to the one described in the previous section. In this kind of model, if the field $\Phi_2$ 
satisfy the bound (\ref{c1}), one could then use the sector
containing $\Phi_2$ and $V$ to break supersymmetry and "uplift" to a Minkowski vacuum a 
supersymmetric AdS minimum for the sector involving $\Phi_1$.

\section{Conclusions}

In this paper we have studied the constraints that can be put on gauge invariant supergravity 
models from the requirement of the existence of a flat and metastable vacuum. Following the 
same strategy as in the previous analysis presented in \cite{grs1,grs2} for the simplest 
supergravity theories involving only chiral multiplets, we have considered the constraints 
implied by the flatness condition implying the vanishing of the cosmological 
constant and by a necessary but not sufficient condition for metastability, derived by looking 
at the particular complex direction in the scalar field space that corresponds to the Goldstino 
direction. These conditions define two algebraic constraints on the chiral and vector auxiliary 
fields $F$ and $D$, and depend on the curvature of the K\"ahler potential, the mass of the 
vector fields and the derivatives of the gauge kinetic function. The major difficulty in solving 
more explicitly these constraints comes from the kinematical and the dynamical relations 
existing between the chiral and vector auxiliary fields. We have presented and followed three 
different methods to derive more explicitly the restrictions imposed by the flatness and stability 
constraints on the parameters defining the theory. These methods are based on the use of respectively 
the dynamical relation, the kinematical relation and a kinematical bound between the $F$ and 
$D$ auxiliary fields, and preserve a decreasing amount of the information contained in the 
original conditions. In this way, we were able to obtain several kinds of necessary conditions, 
which are relevant in different situations.

Our results can be summarized as follows: as expected, the presence of vector multiplets, in 
addition to chiral multiplets, tends to alleviate the constraints with respect to a situations with 
only chiral multiplets. This is mainly due to the fact that the $D$-type auxiliary fields give a positive 
definite contribution to the scalar potential, on the contrary of the $F$-type auxiliary fields, which 
give an indefinite sign contribution. The effect of vector multiplets is maximal when the gauge boson masses 
are comparable to the gravitino mass. When these two mass scales are instead hierarchically different, the 
effect of vector multiplets is small and is encoded in a shift of the components of the Riemann tensor. 
More in general, we found through various types of analyses that the main effect of the vector multiplets 
is essentially to reduce the effective curvature felt by the chiral multiplets, and thereby to make 
the condition of metastability less constraining. On the other hand, the local symmetries associated 
to the vector multiplets and the corresponding Higgs mechanism controlling their spontaneous 
breaking imply that the presence of vector fields does not merely represent an extra generalizing ingredient, but also 
leads to restrictions.

We believe that the general results derived in this paper can be useful in discriminating 
more efficiently potentially viable models among those emerging, for instance, as low-energy
effective descriptions of string models. We leave the exploration of this application for future
work.

\section*{Acknowledgments}

We thank G. Dall'Agata, E. Dudas, R. Rattazzi, M. Serone and A. Uranga for useful discussions. This work has 
been partly supported by the Swiss National Science Foundation and by the European Commission 
under contracts MRTN-CT-2004-005104. We also thank the Theory Division of CERN for hospitality.

\end{document}